\documentclass[
aps,prm,
preprint,
noshowkeys,superscriptaddress]{revtex4-1}
\usepackage{graphicx,
bm,xcolor,microtype,multirow,amscd,amsmath,amssymb,amsfonts,physics,longtable,wrapfig,txfonts}

\usepackage[utf8]{inputenc}
\usepackage[T1]{fontenc}
\usepackage{txfonts}

\usepackage[
	colorlinks=true,
    citecolor=blue,
    breaklinks=true
	]{hyperref}
\begin{document}	

\title{  Universal polarization energies for defects in monolayer, surface and bulk hexagonal boron nitride : A  finite-size fragments $GW$ approach }

\author{David Amblard}
\affiliation{Univ. Grenoble Alpes, CNRS, Inst NEEL, F-38042 Grenoble, France}
\author{Gabriele D'Avino}
\affiliation{Univ. Grenoble Alpes, CNRS, Inst NEEL, F-38042 Grenoble, France}
\author{Ivan Duchemin}
\affiliation{Univ. Grenoble Alpes, CEA, IRIG-MEM-L\_Sim, 38054 Grenoble, France}
\author{Xavier Blase}
\affiliation{Univ. Grenoble Alpes, CNRS, Inst NEEL, F-38042 Grenoble, France}

\begin{abstract}
  We study defect energy levels in hexagonal boron-nitride with varying number of layers  using a fragment many-body $GW$ formalism, taking as examples the paradigmatic carbon-dimer and $C_BV_N$ defects. 
  We show that a single layer can be fragmented in polarizable finite-size areas reproducing faithfully the effect of the dielectric environment, dramatically facilitating the study at the many-body level of point defects in the dilute  limit.
%The small error induced by neglecting the wavefunction overlap between layers in the fragment approach is shown to be largely canceled by calculating polarization energy shifts within the  static Coulomb-hole plus screened exchange approximation to $GW$. 
  The evolution of defect energy levels from the monolayer to a $n$-layer system due to increased screening, labeled polarization energies,  follow a simple $({\Delta P}/n + P_{\infty})$  behavior. The coefficients $\Delta P$ and $P_{\infty}$ are found to be close-to-universal,  
  with opposite signs for holes and electrons,
  characterizing mainly the host and the position of the defect (surface or bulk), but hardly the defect type. 
  % As compared to the monolayer, occupied (unoccupied) defect energy levels are shifted by a close-to-universal polarization energy $P_{\infty} \approx$  0.6~eV ($P_{\infty} \approx$ -0.7~eV) for defects in the bulk.  
  %Such polarization, or screening, effects are    shown to preserve the energy difference between defect occupied levels and the \textit{h}-BN valence band edge. 
  Our results rationalize the evolution of defect energy levels with layers number, allowing to safely extrapolate results obtained for the monolayer  to few-layers, surface  or bulk \textit{h}-BN.    
  The present many-body fragment approach further opens the door to studying disordered  2D layers.  
  %% \textcolor{red}{[169 words, 1182 characters]} 
  \end{abstract}

\maketitle

%%%%%%%%%%%%%%%%%%%%%%%%
\section{Introduction}
\label{sec:intro}

Defects in hexagonal boron-nitride (\textit{h}-BN) have generated much research activity both at the experimental 
 \cite{Katzir75,Era81,Silly07,Museur08,Meuret_2015,Wong_2015,Tran_2016,Bourrellier_2016,Vuong16,Jungwirth_2016,Gottscholl20,Hayee20,Mendelson21}
 %Wu04,Du15,Vokhmintsev_2019,
and theoretical
\cite{Attaccalite_2011,Attaccalite_2013,Berseneva13,Wu_2017,Huang12,Tawfik_2017,Cheng_2017,Jungwirth17,Reimers_2018,Abdi_2018,Weston_2018,Smart_2018,Sajid_2018,Turiansky_2019,Korona_2019,Mackoit_2019,Wang_2020,Chen_2021,Reimers_20,Sajid_2020,Bhang_2021,Jara2021,Winter_2021,Gao_2021,Libbi_2022}
%%Wang_2019,Linderalv_2021,
level to better understand their unique photoluminescent properties. 
The large gap of the host \textit{h}-BN allows a variety of localized occupied and unoccupied defect states  controlling  the absorption and emission spectrum across the visible range. 
While the optical properties of defects can be used as a way to unravel their chemical nature, thanks to the comparison between theoretical and experimental optical emission energies and lineshapes, 
much less is known experimentally about their electronic energy levels as measured by photoemission. 
Still, optical properties are strongly related to an accurate description of defect electronic properties, together with the strength of the excitonic electron-hole screened interaction. 
Further, the position of the defect energy levels in the host gap, in conjunction with structural reorganization energies,  allows to predict its charge state as a function of the chemical potential. 

The exploration of the electronic properties of defects in the dilute limit using  \textit{ab initio}  simulations  remains a difficult task. The most common periodic boundary condition (PBC) calculations, with defects repeated periodically, requires rather large unit cells to minimize spurious defect-defect interactions. This is all the more a difficulty in the case of charged defects for which  Coulomb truncation techniques must be used to annihilate long-range Coulomb interactions between periodic images \cite{Freysoldt14}. Further, at the efficient density-functional theory (DFT) Kohn-Sham level, the position of the electronic energy levels is known to be strongly sensitive to the choice of the exchange-correlation functional, limiting the predictive character of the related calculations. 
%This dictates the use of some tuning scheme for  the amount of exact exchange, distinguishing possibly short and long-range amounts for range-separated hybrid functionals. 
In the absence of experimental photoemission data characterizing defect levels, this may lead to difficulties on the best strategy to adopt. 
Recently, a scheme based on enforcing Koopman's like conditions \cite{Miceli_2018} illustrated the importance of changing the amount of exact exchange at the generalized Kohn-Sham DFT level as a function of the number of  \textit{h}-BN layers~\cite{Smart_2018}.

%% Attaccalite_2011    defect in monolayers
%% Berseneva13         defects in monolayers + BN monolayer and bulk with various G0W0, GW0, GW schemes ..
%% Wu_2017                GW monolayers
%% Duc_21                    GW monolayers
%% Chen_2021     GW/BSE monolayers
%% Winter_2021 monolayers
%% GW on 3 layer systems max,  

 In the family of techniques going beyond Kohn-Sham DFT for the study of electronic energy levels, the many-body Green's function $GW$ perturbation theory~\cite{Hed65,Str80,Hyb86,Farid88,God88} stands as a now common approach offering a reasonable compromise between accuracy and computer cost.  Following pioneering studies for defects in 3D systems  \cite{Surh_1995,Hedstrom_2006,Rinke_2009,Martin-Samos_2010,Jain_2011,Wei_2013}, $GW$ calculations for defects  in cubic  \cite{Tararan_2018} and hexagonal     \cite{Attaccalite_2011,Attaccalite_2013,Berseneva13,Wu_2017,Smart_2018,Chen_2021,Winter_2021,Gao_2021,Libbi_2022} BN  have been conducted in a few groups. Of specific interest for the present study, such a formalism  accounts for the interaction of an added charge with the N-electron system, properly mimicking a photoemission experiment. This interaction is constructed using the calculated linear-response electronic susceptibility    that adapts to the dielectric environment. As such, the related screened Coulomb potential  $W$ fully accounts for the change in dielectric conditions from the monolayer to the few-layers or bulk limit. 
 
 As a caveat of properly accounting for charging effects, $GW$ calculations with periodic boundary conditions also encounters difficulties associated with spurious long-range Coulomb interactions between images in the case of periodic boundary calculations. As recently analyzed in the case of defected \textit{h}-BN systems \cite{Wu_2017},  2D  $GW$ calculations converge slowly, both with the  amount of vacuum between repeated images and with  the in-plane unit-cell size.  
 While Coulomb truncation techniques can been used to avoid spurious interlayer interactions for the monolayer limit \cite{Rozzi_2006,Huser_2013,Wu_2017},   
eliminating  interactions between periodically repeated defects within a layer, without affecting  the in-plane  screening  by the \textit{h}-BN host, stands as a difficult challenge. 
 In the  bulk limit, the necessity to include a sufficient number of pristine layers in the c-axis direction further increases computational cost. As a matter of fact, $GW$ calculations for defects in \textit{h}-BN were conducted essentially for monolayers \cite{Attaccalite_2011,Berseneva13,Wu_2017,Smart_2018,Chen_2021,Winter_2021,Libbi_2022}, with one study of defected 3-layer systems \cite{Smart_2018} and one bulk study with 2-layers per unit-cell \cite{Gao_2021}.
 
% In  a recent study, many-body $GW$ calculations were performed for  the carbon-dimer defect in a finite-size  \textit{h}-BN cluster geometry \cite{Winter_2021} . Varying the size of the clusters containing the defect, extrapolation to the monolayer single-defect dilute limit could be obtained as an alternative to periodic boundary calculations. It was shown in particular that electronic and optical energies scale as 1/$N_{at}$ and 1/$N_{at}^{3/2}$, respectively, with $N_{at}$ the number of atoms in the 2D cluster. 

We  propose in the present work an alternative finite-size cluster approach, performing  $GW$ calculations of defect   energy levels in \textit{h}-BN, reaching the  monolayer,   few-layers, surface and bulk dilute limit conditions. Defected and pristine flakes with increasing lateral size are stacked, reaching sizes large enough to safely extrapolate to the infinite  limit for a given number of layers. We show that convergence with respect to in-plane dimensions can be dramatically facilitated by adopting a fragment scheme  where each layer is reconstructed by patching large \textit{h}-BN clusters,  neglecting  wavefunctions overlap between fragments on the same or neighboring layers. 
%This dramatically reduces the cost of constructing the electronic susceptibility, while reproducing faithfully long-range polarization effects. 
This allows to study at the many-body level systems containing several thousand atoms, reaching sizes  large enough to reliably extrapolate to the infinite 2D or 3D limits. Defect  energy level variations from the monolayer to a $n$-layer system  follow a simple $({\Delta P}/n + P_{\infty})$  polarization energy behavior,  where the rate and asymptotic coefficients $\Delta P$ and $P_{\infty}$ characterize mainly the host and the position of the defect (surface or bulk), but hardly the defect type. 
%Delocalized \textit{h}-BN states at the valence band maximum are characterized by similar polarization energies, allowing to follow the evolution of defect levels with respect both to the vacuum level and the valence band maximum upon changing the environmental screening conditions.   
Our study rationalizes the evolution of defect energy levels as a function of the number of layers.
As a result, data  obtained for the monolayer,  or very-few layer systems, can be easily extrapolated to the surface or bulk limit without the need to perform additional many-body calculations.

% Experimental few layers hBN with defects \cite{Mendelson_2019} including monolayer \cite{Jin_2009}.
% $GW$ calculations for defects \cite{Hedstrom_2006,Rinke_2009,Martin-Samos_2010,Jain_2011,Wei_2013,Malashevich_2014,Wu_2017,Smart_2018,Chen_2021}
% $GW$ calculations of defects in monolayer and few layers h-BN using periodic boundary conditions  \cite{Attaccalite_2011,Wu_2017,Smart_2018} or extrapolation from finite-size cluster calculations \cite{Winter_2021}. 

\section{ Methodology }

We briefly outline in this Section the methodology adopted in the present study. More details on the $GW$ formalism \cite{Hed65,Str80,Hyb86,Farid88,God88} can be found in the thorough reviews written on the subject  \cite{Ary98,Farid99,Onida02,Ping_2013,Golze_2019,ReiningBook}. Let's recall for the present purpose that the poles of the one-body Green's function $G$ represent by definition proper electron addition $\; E_n[N+1]-E_0[N] \;$ or removal $\; E_0[N]-E_n[N-1] \;$   energies, with $E_n[N \pm 1]$ the total energy of the ($N \pm 1$) electron system in its $n$-th excited state, and $E_0[N]$ the ground-state energy of the N-electron system. As such, the poles of the Green's function  account for the interaction of an added charge with the rest of the system, including with a polarizable environment through  its contribution to the screened Coulomb potential $W$. While the price of calculating the screened Coulomb potential for large multilayer systems in the dilute defect limit is prohibitively expansive, the fragment approach that neglects wavefunction overlaps between layers, and further  between subsections of a same layer,  leads to a dramatic cost reduction with small induced errors. 
 
\subsection{ Fragment \textit{GW} calculations }

We restrict ourselves to define   the quantities of interest in the present paper, namely the $GW$ self-energy and the screened Coulomb potential $W$. The $GW$ self-energy  represents an approximation to the dynamical exchange-correlation operator that appears in the equation-of-motion for the time-ordered one-body Green's function $G$, with: 
\begin{align}
 \Sigma^{GW}({\bf r},{\bf r}' ; E) &= \frac{i}{2\pi} \int d {\omega}  \; e^{i \omega \eta } 
     G({\bf r},{\bf r}' ;  E+\omega) W({\bf r},{\bf r}' ; \omega) \label{eqn:sigma} \\
      W({\bf r},{\bf r}' ; \omega)  
    &=    v({\bf r},{\bf r}') +   \int d{\bf r}_1 d{\bf r}_2 \;  v({\bf r},{\bf r}_1) \chi_0({\bf r}_1,{\bf r}_2 ; \omega)  W({\bf r}_2,{\bf r}' ; \omega) \label{eqn:Wscr} \\
 \chi_0({\bf r},{\bf r}' ; \omega)  
    &=   \sum_{mn}  (f_m - f_n)  \frac{   \phi_n^*({\bf r}) \phi_m({\bf r})        \phi_m^*({\bf r}')  \phi_n({\bf r}')   }{ \omega - ( \varepsilon_m - \varepsilon_n) + i \eta}    \label{eqn:chi0}
\end{align}
where $\chi_0$ is the independent-electron  susceptibility and $v$ the bare Coulomb potential.  The terms $\lbrace f_{n/m} \rbrace$ are level occupation   factors and $\eta$ a  positive infinitesimal. In practice, the needed input Green's function $G$ and susceptibility $\chi_0$ are calculated with  input $\lbrace \varepsilon_n^{KS}, \phi_n^{KS} \rbrace$ Kohn-Sham eigenstates, e.g.:
$$
G^{KS}({\bf r},{\bf r}' ; \omega) =
\sum_n \frac{ \phi_n^{KS}({\bf r})  [\phi_n^{KS}({\bf r}')]^{*}  }{ \omega - \varepsilon_n^{KS} + i\eta \times \text{sgn}(\varepsilon_n^{KS} - \mu) } 
$$
with $\mu$ the Fermi energy. The knowledge of the self-energy operator allows to calculate the $GW$ Green's function and energy levels by replacing the DFT exchange-correlation potential contribution by the $GW$ self-energy operator calculated at the targeted $GW$ energy, namely:
$$
G(\omega) = G^{KS}(\omega) + G^{KS}(\omega) \left[ \Sigma^{GW}(\omega) - V^{XC}_{DFT} \right] G(\omega)
$$
and  for the energy levels:
$$
\varepsilon_n^{GW} = \varepsilon_n^{KS} + \langle \phi_n^{KS} | \Sigma^{GW}( \varepsilon_n^{GW}  ) - V^{XC}_{DFT} | \phi_n^{KS} \rangle
$$
Such a scheme, where the self-energy is built from the Kohn-Sham input eigenstates, is labeled the single-shot, or $G_0W_0$, approach, as compared to self-consistent techniques where corrected eigenvalues, possibly eigenstates, are reinjected to calculate $G$ and $W$.

In the fragment $GW$ approach, one assumes that the system is partitioned in subsystems (labeled also fragments or clusters below) with weakly overlapping wavefunctions. In that limit, the independent-electron susceptibility $\chi_0$ (Eq.~\ref{eqn:chi0}) has hardly any  contribution from orbitals $\phi_n$ and   $\phi_m$ belonging to different fragments. The susceptibility $\chi_0$ is then diagonal by blocks,   each block built from the susceptibility of one fragment in the isolated (gas phase) limit. This represents a considerable saving since in the fragment limit  the cost of calculating the overall $\chi_0$ does not grow quartically with  system size, but only
linearly. If the fragments are identical, just translations of the same cluster  with identical susceptibility blocks,  calculating the overall $\chi_0$ is independent of the system size. Such a fragment $GW$ approach was used in particular in the study of $\pi$-conjugated organic crystals, with weak interactions between molecular units, \cite{Fujita_2018,Tolle_2021}  nanotube bundles \cite{Spataru_2013} or  layered 2D systems bound by weak van der Walls interactions, defining the field of 2D-genomics  \cite{Andersen_2015,Winther_2017,Druppel_2018,Xuan_2019}.    Further, such a partitioning of the independent-electron susceptibility served as the basis for combining the $GW$ and Bethe-Salpeter \cite{Sha66,Han79,Str82} formalisms with semi-empirical continuous or discrete   models of polarizable environments  \cite{Baumeier_2014,Li_2016,Duchemin_2016,Duchemin_2018,Wehner_2018}.

\begin{figure}[t]
       \includegraphics[width=8cm]{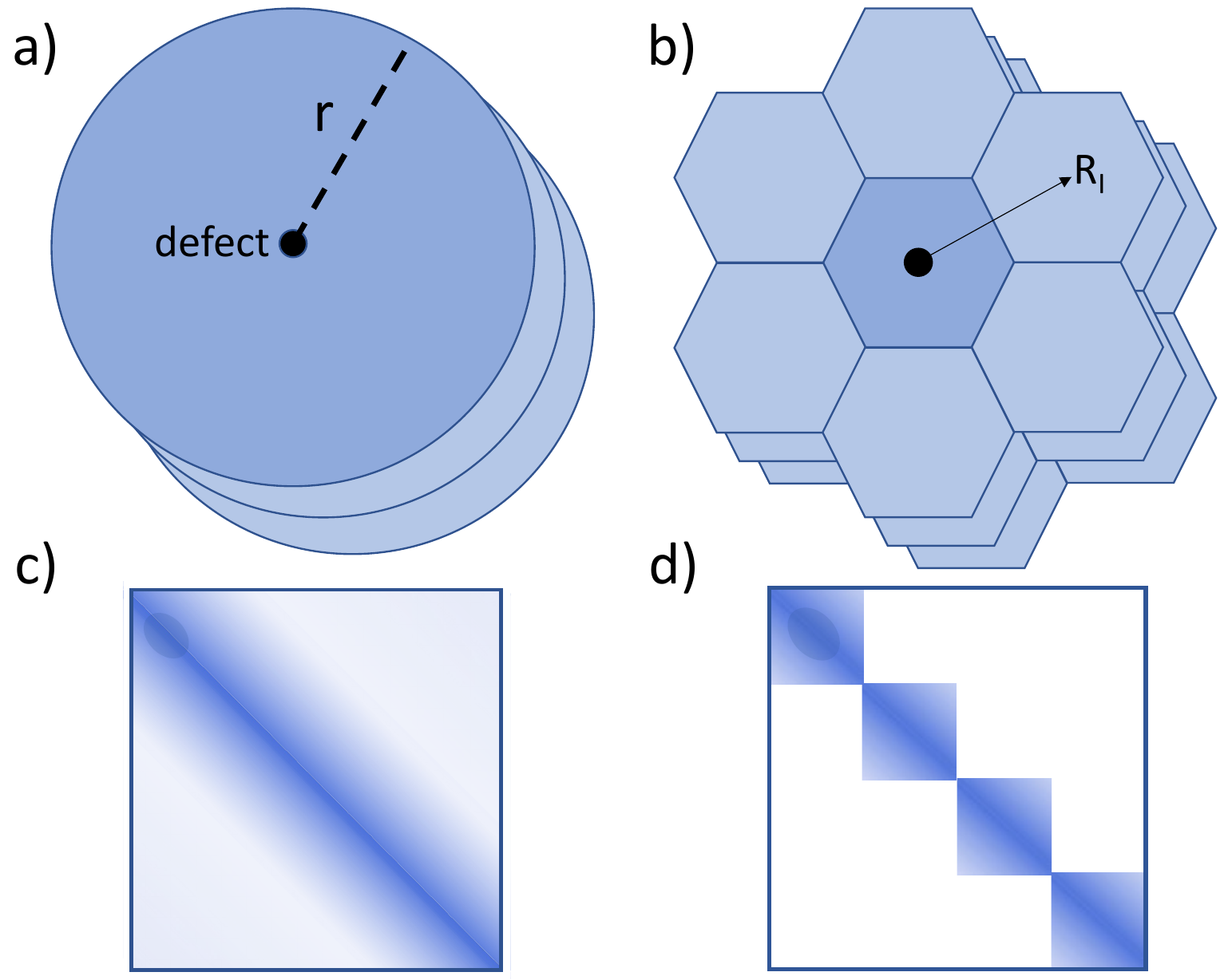}
        \centering
        \caption{ Symbolic representation of a finite-size multilayers system. The defect is represented here by the black dot at the center of the surface layer or central fragment in darker blue.  In (a), convergence to the infinite-layer limit can be achieved by increasing the radius r with a large computational cost increase, while in (b) the same limit is reached by adding shells of first, second, etc. nearest neighbor domains, namely undefected fragments translated by a set of $\lbrace {\bf R}_I \rbrace$ vectors. The corresponding independent-electron susceptibility matrices are represented in c) and d) respectively. 
       %% In c), the fact that on general grounds the independent-electron susceptibility operator $\chi_0({\bf r},{\bf r}'; \omega)$ decays for $ | {\bf r}-{\bf r}' | \rightarrow +\infty$ is rendered by a diagonal-dominant matrix with a colormap representation. 
       %% Such a property is not exploited in the present study but may be used to better understand the success of the fragmentation scheme. 
        }
        \label{fig:scheme1}
\end{figure}

In the present study, we not only neglect the wavefunction overlap between neighboring layers, as already done in several studies  \cite{Andersen_2015,Winther_2017,Druppel_2018,Xuan_2019}, but we further fragment individual monolayers in domains. This is done by ``reconstructing" monolayers by patching  \textit{h}-BN clusters as symbolically represented in Fig.~\ref{fig:scheme1}. Writing the independent-electron susceptibility as a sum over fragments (indexed by $I$) in matrix notations, with ($I=0$) the defected fragment (or cluster), one writes: 
\begin{align*}
  \chi_0({\bf r},{\bf r}'; \omega) &= 
  \chi_0^{(I=0)}({\bf r},{\bf r}'; \omega) + \sum_{I>0} \chi_0^{(I)}({\bf r},{\bf r}'; \omega) \\
  &= \chi_0^{(defect)}({\bf r},{\bf r}'; \omega) + \sum_{I>0} \chi_0^{(pristine)}({\bf r}-{\bf R}_I,{\bf r}' - {\bf R}_I; \omega) 
\end{align*}
where $\chi_0^{(pristine)}$ is the susceptibility of an undefected (pristine) \textit{h}-BN cluster and the $\lbrace {\bf R}_I \rbrace$ are the translations needed to reconstruct all layers with \textit{h}-BN fragments. 
Such an approach only requires the susceptibility $\chi_0^{(defect)}({\bf r},{\bf r}'; \omega)$ of an isolated defected \textit{h}-BN cluster and the susceptibility 
$ \chi_0^{(pristine)}({\bf r},{\bf r}' ; \omega)$  of an undefected \textit{h}-BN cluster that will be ``translated" to fill up the full system $\chi_0({\bf r},{\bf r}'; \omega)$ matrix (see Fig.~\ref{fig:scheme1}d).  For the defected layer, the central fragment contains the defect, surrounded by shells of first, second, etc. nearest-neighbor pristine \textit{h}-BN clusters. In practice, each layer contains up to 4-th neighbor  fragments, amounting to 57 clusters  containing typically 86 to 138 atoms each, reaching lateral sizes large enough to allow extrapolation to the infinite layer size limit for a given number of layers.   
The accuracy of this scheme will be carefully validated below.

%%% Schema1  1  %%%

While the non-interacting  $\chi_0$ blocks are decoupled, it is important to keep in mind that all fragments interact in the construction of the screened Coulomb potential $W$. With the notations defined above,  the Dyson-like equation ~\ref{eqn:Wscr}  allows to write the screened Coulomb potential in block notations as: 
\begin{equation*}
W = v +  v   \Big[ \sum_I \chi_0^{I}   + \sum_{IJ} \chi_0^{I}   v_{IJ}   \chi_0^{J} 
 + \sum_{IJK} \chi_0^{I}   v_{IJ}   \chi_0^{J}   v_{JK}   \chi_0^{K} + \cdots \Big]   v
\end{equation*}
where the $v_{IJ}$ are the bare Coulomb matrix elements between  fragments (I) and (J). 
As such, the screened Coulomb potential includes all order interactions between fragments via the bare Coulomb potential.

We conclude by noting that while the cost of building the independent-electron susceptibility, the most expensive step in a $GW$ calculation, is dramatically reduced in the fragment approach,  inverting the Dyson equation for $W$ remains to be done (Eq.~\ref{eqn:Wscr}), with a cubic scaling cost with system size.  However, systems with hundreds of domains, adding up to thousands of atoms, can be treated fully \textit{ab initio} as discussed here below.

\subsection{ The static COHSEX approximation for environmental screening}

To further reduce the cost of inverting the Dyson equation for $W(\omega)$ at each frequency, we further explore  the merit of a popular approximation to the full $GW$ dynamical self-energy operator labelled  the static Coulomb-hole (COH) plus screened exchange (SEX) approximation. The COHSEX  terminology  describes  an exact partitioning of the self-energy, gathering in the SEX/COH terms the contributions from the poles of $G$/$W$ \cite{Hedin65,Hybertsen_1986,Far88}.  The static approximation takes to zero all energy differences $(\varepsilon_n - \omega)$ in the COHSEX terms, where $\varepsilon_n$ covers the spectrum of electronic energy levels and $\omega$ is the frequency at which the self-energy is to be calculated, yielding:

\begin{align*}
%\label{eq:static_SEX}
\Sigma^{SEX}_{static}({\bf r},{\bf r}';0)  &=  - \sum_n^{occ} \phi_n({\bf{r}}) \phi_n^*({\bf{r}'})  W({\bf{r}},{\bf{r}}'; \omega=0) \\
\Sigma^{COH}_{static}({\bf r},{\bf r}';0)  &=   {1 \over 2} \sum_n \phi_n({\bf{r}}) \phi_n^*({\bf{r}}')
(W-v)({\bf{r}},{\bf{r}'}; 0) \nonumber \\
                 &=   \frac{1}{2} \delta({\bf{r}}-{\bf{r}'})  (W-v)({\bf{r}},{\bf{r}}'; \omega=0).
\end{align*}

This static approximation offers the advantage that the screened Coulomb potential only needs to be calculated at the low-frequency limit $(\omega \rightarrow 0)$, greatly reducing computer cost. In the present analytic continuation implementation of $GW$ \cite{Duc20,Duc21}, for which the susceptibility needs to be calculated at typically 12 frequencies along the imaginary axis, the static COHSEX approximation is thus an order-of-magnitude cheaper than the full $GW$ calculation. 

Such an approximation was shown to yield significantly too large gaps \cite{Hybertsen_1986} but has been central to most studies implementing the fragment many-body techniques \cite{Neaton_2006,Tolle_2021} or the combination of many-body techniques with models of dielectric environment   \cite{Duchemin_2016,Duchemin_2018}. In such studies, environmental corrections are calculated within the static COHSEX approximation in the form of a difference, namely that of the static COHSEX values of an energy level with and without the environment. In practice, this means that the $GW$ energy levels for a defected fragment surrounded (embedded) by undefected \textit{h}-BN flakes is calculated as:
\begin{equation*}
\varepsilon_n^{eGW} \simeq \varepsilon_n^{gGW} + ( \varepsilon_n^{e\text{COHSEX}} - \varepsilon_n^{g\text{COHSEX}} )
\end{equation*}
where $eGW$ and  $e$COHSEX  mean embedded $GW$ and COHSEX calculations,  while $gGW$ and $g$COHSEX  point to gas phase, namely isolated  defected fragment calculations.

The difference $\Delta_n^{GW} = ( \varepsilon_n^{eGW} - \varepsilon_n^{gGW} ) $ and its approximation
$\Delta_n^{\text{COHSEX}} = ( \varepsilon_n^{e\text{COHSEX}} - \varepsilon_n^{g\text{COHSEX}} ) $ that characterizes the evolution of   electronic energy levels from the monolayer to the multilayer cases (few-layers, surface, bulk) is the main targeted observable of the present study. Such an energy shift may be decomposed into two contributions : (a) the shift of the Kohn-Sham energies $\Delta_n^{KS} = ( \varepsilon_n^{e\text{KS}} - \varepsilon_n^{g\text{KS}} )$ and (b) the shift due to screening effects
at the ${\Delta}GW$ or $\Delta$COHSEX level. The quantity $\Delta_n^{KS}$ accounts for hybridization, confinement, or electrostatic effects in the ground-state, while the second  occurs as a response to an electronic excitation (e.g. photoemission) on the defect. We will  label  polarization energy this second contribution that can be defined as the difference $P_n = \Delta_n^{GW} - \Delta_n^{KS}$. While   $\Delta_n^{KS}$ can shift up or down energy levels, the second effect always stabilizes holes and electrons,   namely pushing occupied levels towards the vacuum level, while on the contrary empty levels go down in energy.

%Physically, the use of the static COHSEX approximation implies that only the low frequency limit (in the optical range) of the environmental dielectric response is considered.  \textcolor{red}{ Such an adiabatic approximation, which assumes that the environment reacts instantaneously to an excitation in the central system of interest, may be questionable in cases where the gap in the environment electronic excitation spectrum is not clearly larger than that of the central object of interest \cite{Huu2020}.   }
Physically, the use of the static COHSEX approximation implies the neglect of the frequency dependence of the dielectric response, assuming that the environment reacts instantaneously to an excitation in the central system of interest, with the susceptibility obtained in the low-frequency limit.  This approximation may be questionable in cases where the gap in the environment electronic excitation spectrum is not clearly larger than that of the central object of interest.
We will thus for validation compare $\Delta GW$ and static $\Delta$COHSEX  polarization energies.

%%%%%%%%%%%%%%%%%%%%%%%%
\subsection{Technical details}
\label{sec:technical}

Our DFT calculations are performed with the ORCA code \cite{orca} that is used to obtain relaxed geometries and generate the Kohn-Sham eigenstates used as an input for many-body $GW$ corrections. Such many-body calculations are performed with the BeDeft (Beyond-DFT) package \cite{Duc20,Duc21}, an evolution of the  {\sc{Fiesta}} code  \cite{Jac15a,Li16,Jac17,Duc18}, that implements the $GW$ formalism within a Gaussian basis formalism. The dynamical self-energy is obtained using an improved analytic continuation scheme \cite{Duc20,Duc21} that does not require any plasmon-pole approximation. All occupied, including core states, and unoccupied energy levels are included in the construction of the susceptibility and self-energy. Coulomb integrals are calculated  using standard Coulomb-fitting resolution of the identity \cite{Whitten73,Ren_2012,Duchemin17}. At that stage, we emphasize that the present all-electron finite-size calculations with Gaussian basis sets, not relying on the Bloch theorem nor on the use of pseudopotentials, provide energy levels directly referenced to the vacuum level.

Structural relaxations are performed at the triple-zeta plus polarization 6-311G(d) level \cite{Krishnan80}. We adopt the same basis for the calculation of energy differences at the $GW$ and static COHSEX level. While absolute $GW$ energies are not converged with such a basis set,  energy differences between the defected monolayer and corresponding multilayer systems are very well captured.
%  as demonstrated by comparison with the much larger correlation-consistent cc-pVTZ basis set. \cite{Dunning89} 
Such energy differences involve long-range polarization effects that converge very quickly with basis size, as further demonstrated below in our discussion on the basis-set dependence of the fragments dipolar response. 
%To obtain reference energies with respect to the vacuum level, $GW$ calculations on the defected BN86 flakes are performed at the cc-pVXZ level, where X ranges from double, triple, to quadrupole level, followed by an extrapolation to the complete basis set limit (see Supplemental Material \cite{supplemental}). 

Our $GW$ calculations are performed at the non-self consistent $G_0W_0$ level starting with a PBEh($\alpha$) functional \cite{PBE0} with 40$\%$ of exact exchange ($\alpha=0.4$). In the process of calculating  polarization energies, input Kohn-Sham eigenstates enter indirectly through the construction of the $\chi_0$ Kohn-Sham independent-electrons susceptibility. Since we are interested in exploring the impact of the dielectric response of added layers on the defect, we adopt an $\alpha$ value close to the optimal one ($\alpha=0.409$) found in Ref.~\citenum{Smart_2018} for reproducing the monolayer long-range dielectric properties in an optimally tuned-functional approach. We emphasize however that Kohn-Sham eigenstates are used in the present case as an input starting guess to build the dielectric function and the $GW$ self-energy. This $GW$ correction  dramatically reduces the impact of the chosen DFT functional on the final quasiparticle energies. Anticipating on the upcoming results, the shift in energy for the  carbon-dimer defect levels from the monolayer to the \textit{h}-BN bulk changes   by no more than  10$\%$ taking PBE Kohn-Sham eigenstates instead of PBEh(0.4) ones as starting points, even though the related Kohn-Sham gaps differ by $\sim$3 eV   (monolayer limit). 

%% Gap C2BN86   PBE = 3.673 eV   and in PBEh(0.4) = 6.66 eVs
 
We now address an important aspect of our calculations pertaining to the idea of multipole expansions. In the standard  Coulomb-fitting  resolution-of-the-identity (RI-V) technique we adopt  \cite{Whitten73,Ren_2012,Duchemin17},   orbital products $\phi_m({\bf r})\phi_n({\bf r})$ appearing in the construction of the independent electron susceptibility (Eq.~\ref{eqn:chi0}) are expressed over an auxiliary basis $\lbrace P_{\mu} \rbrace$ of Gaussian orbitals:
\begin{align*}
\phi_m({\bf r})\phi_n({\bf r}) &= \sum_{\mu} C_{\mu}^{nm} P_{\mu}({\bf r})  \\
C_{\mu}^{nm} & \stackrel{RI-V}{=}  \sum_{\nu} [V^{-1}]_{\mu\nu} (\phi_m\phi_n | P_{\nu} )  
\end{align*}
where $V_{\mu\nu}$ are Coulomb matrix elements in the auxiliary basis and $( \cdot | \cdot )$ denotes a  Coulomb integral. Such $\phi_m({\bf r})\phi_n({\bf r})$ products, and their representative auxiliary basis set, enter in the description of the charge rearrangements induced by a perturbation in the system.  In finite size systems, molecular orbitals can be taken to be real. 
The independent-electron susceptibility is calculated in the auxiliary basis with:
\begin{align*}
 \chi_0( {\bf r} , {\bf r}' ; i \omega) &= \sum_{\mu\nu}   [\chi_0(i \omega)]_{\mu\nu}   P_{\mu}({\bf r}) P_{\nu}({\bf r}') \\
 [\chi_0( i \omega)]_{\mu\nu}    
     &=   \sum_{mn}  \frac{ (f_m - f_n)  C_{\mu}^{nm} C_{\nu}^{nm}   }{ i \omega - ( \varepsilon_m - \varepsilon_n) }   \label{eqn:chi0}
\end{align*}
Such  auxiliary basis sets, that span  the ``product-space" of Kohn-Sham orbitals,  have been optimized for each standard Gaussian Kohn-Sham basis and we use the universal Coulomb fitting auxiliary basis \cite{Weigend06} in conjunction with the 6-311G(d) basis set. 
It is in such an auxiliary basis representation that the Dyson equation for the screened Coulomb potential $W$ is inverted or, equivalently, the interacting susceptibility   $\chi$:   
\begin{align*}
    \chi({\bf r},{\bf r}'; \omega) &= \chi_0({\bf r},{\bf r}'; \omega)   
     + \int d{\bf r}_1 d{\bf r}_2 \; \chi_0({\bf r},{\bf r}_1; \omega) v({\bf r}_1,{\bf r}_2) \chi({\bf r}_2,{\bf r}'; \omega)  
\end{align*}
Reducing the size of the auxiliary basis  can quickly degrade the quality of correlated calculations. We have observed however that the polarizability tensor $\lbrace \alpha_{ij} \rbrace$ of a given fragment:
$$
\alpha_{ij}^{(I)}  = - \int d{\bf r} d{\bf r}' \; r_i \cdot  \chi^{(I)}({\bf r},{\bf r}'; \omega) \cdot  r^{\prime}_j
$$
with $\chi^{(I)}$ the interacting susceptibility of fragment ($I$) in the gas phase, is hardly affected by reducing significantly the auxiliary basis set. In the case of the \textit{h}-BN fragments considered in this study, removing the  (\textit{d,f,g})-character orbitals and the core (\textit{s,p}) ones from the auxiliary $\lbrace P_{\mu} \rbrace$ basis set hardly changes the fragment dipolar polarizability, with error well below the percent. We use such  reduced auxiliary basis sets to expand the microscopic susceptibility $\chi_0^{(I)}$  of fragments that are beyond the first-nearest-neighbors   of the central defected fragment of interest. As a result, the calculated polarization energies from the monolayer to the multilayer cases do  not change by more than a very few meVs. We interpret this favorable behavior by emphasizing that for long-range interactions,  induced dipoles are the major contribution to the reaction field and polarization energy. As such, an approximation that preserves the calculated dipolar tensor of remote fragments may be very accurate as we observe. We do not attempt to reduce the basis of nearest neighbor fragments. 
%Such a  property allows to study at the many-body level systems containing more than a hundred fragments, each fragment containing 86 to 138 atoms. 

\subsection{Geometries}

%%% FIG  1  %%%
\begin{figure}[h] 
	\includegraphics[width=8.6cm]{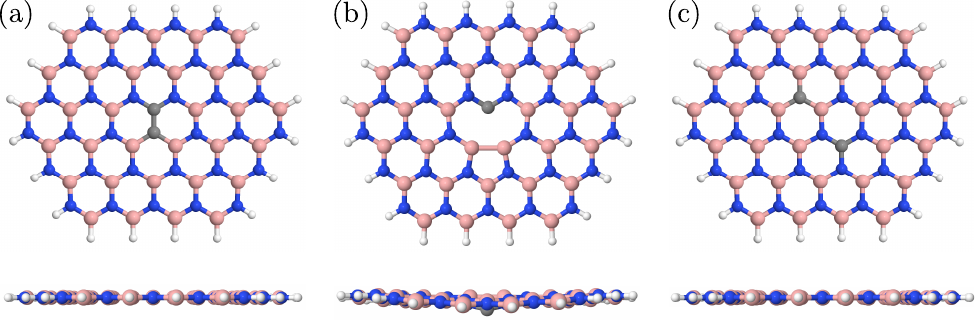}
	\centering
	\caption{ Ball-and-stick representation of the (a) carbon-dimer (CC) defect, (b) the $C_BV_N$ defect and (c) the carbon-dimer defect in the 4-th nearest neighbor configuration labeled CC-$\sqrt{7}$ \cite{Auburger_2021}. Defects are located   at the center of a BN86 flake (or fragment). Edge atoms are passivated by hydrogen. The CC defect in the BN138, BN202 and BN278 fragments are reproduced in the SM~\cite{supplemental}. In the $C_BV_N$ defect, the carbon atom stands $\sim$0.5~\AA\  below the average plane defining the defected layer (see text).  }
	\label{fig:geometries}
\end{figure} 

We study \textit{h}-BN clusters  containing 86, 138, 202 and 278 atoms with passivating hydrogens at the edge. Such systems will be named BN86, BN138, etc. In the case of defected structures, the defect is located at the center of the cluster. We represent in Fig.~\ref{fig:geometries}(a,b,c) the paradigmatic neutral carbon-dimer (labeled CC) and $C_BV_N$ defects, together with the  so-called CC-$\sqrt{7}$ carbon-dimer variant \cite{Auburger_2021}.
In the following, we will use e.g. the notation CC@BN86 for a CC defect at the center of a BN86 flake or cluster. 
The carbon-dimer defect has been proposed \cite{Mackoit_2019, Winter_2021}  to be a likely candidate for the  frequently observed 4.1 eV emission line, while the $C_BV_N$ defect has been associated with the $\sim$2  eV emission line \cite{Sajid_2020b,Sajid_2020}. The CC-$\sqrt{7}$ carbon-dimer defect, with carbon atoms in 4-th nearest neighbor position, is less stable than the standard (nearest-neighbor) CC carbon-dimer defect \cite{Auburger_2021}. However   it exhibits a larger spatial extension and a gap significantly smaller  (by $\sim$2.2 eV at the $GW$ level for the monolayer)  than the standard carbon-dimer defect, allowing to show below that polarization (screening) effects  are nearly independent of the geometrical and electronic properties of the defect of interest. All structures are relaxed at the PBE0 6-311G(d) level. 

Except for the non-planar $C_BV_N$ defect, we do not attempt to relax multilayer systems.  In the dilute  limit, the energetics around the defect is not expected to govern the stacking properties of the \textit{h}-BN layers. As a test, in the case of the carbon-dimer defect located in a BN86 fragment, we prepare a bilayer by adding an undefected BN86 fragment in a AA'  stacking geometry with the experimental 3.33~\AA\ interlayer spacing \cite{Solozhenko1995,Paszkowicz2002}. Structural relaxation including D3 dispersion forces \cite{Grimme2010} preserves the  AA'  stacking geometry, with a slightly reduced 3.25~\AA\ interlayer spacing in the relaxed bilayer. As such, even in the rather high-doping limit, planarity and stacking are preserved for the carbon-dimer defects. In what follows, we will thus relax individually the layers and stack them in an $AA'$ fashion with the experimental 3.33~\AA\ interlayer spacing. 

The case of the $C_BV_N$ defect is more complicated since the relaxation of a single-layer defected fragment leads to a highly non-planar system (see SM~\cite{supplemental}). This is due presumably to the  tensile strain induced by forming a B-B bond across the missing N atom. However, adding a second undefected fragment in AA' stacking and relaxing with D3 dispersion corrections allows to dramatically restore planarity thanks to layer-layer interaction. Nevertheless, the C-atom goes towards the neighbouring layer with an out-of-plane deviation of about 0.55~\AA. The same result is obtained in a trilayer system where the defect is sandwiched between two undefected layers, the C atoms remaining 0.52~\AA\ out of plane despite the restored symmetry of having one layer on each side of the defect. In what follows, we thus adopt the bilayer or trilayer geometry, for studying the surface and bulk limits, and add subsequent layers in an AA' stacking geometry and a 3.33~\AA\ distance with respect to the nearest undefected layer. 

%%%%%%%%%%%%%%%%%%%%%%%%
\section{Fragmenting the monolayer}
\label{sec:fragmono}

%%% FIG  2 %%%
\begin{figure}[t] 
	\includegraphics[width=8cm]{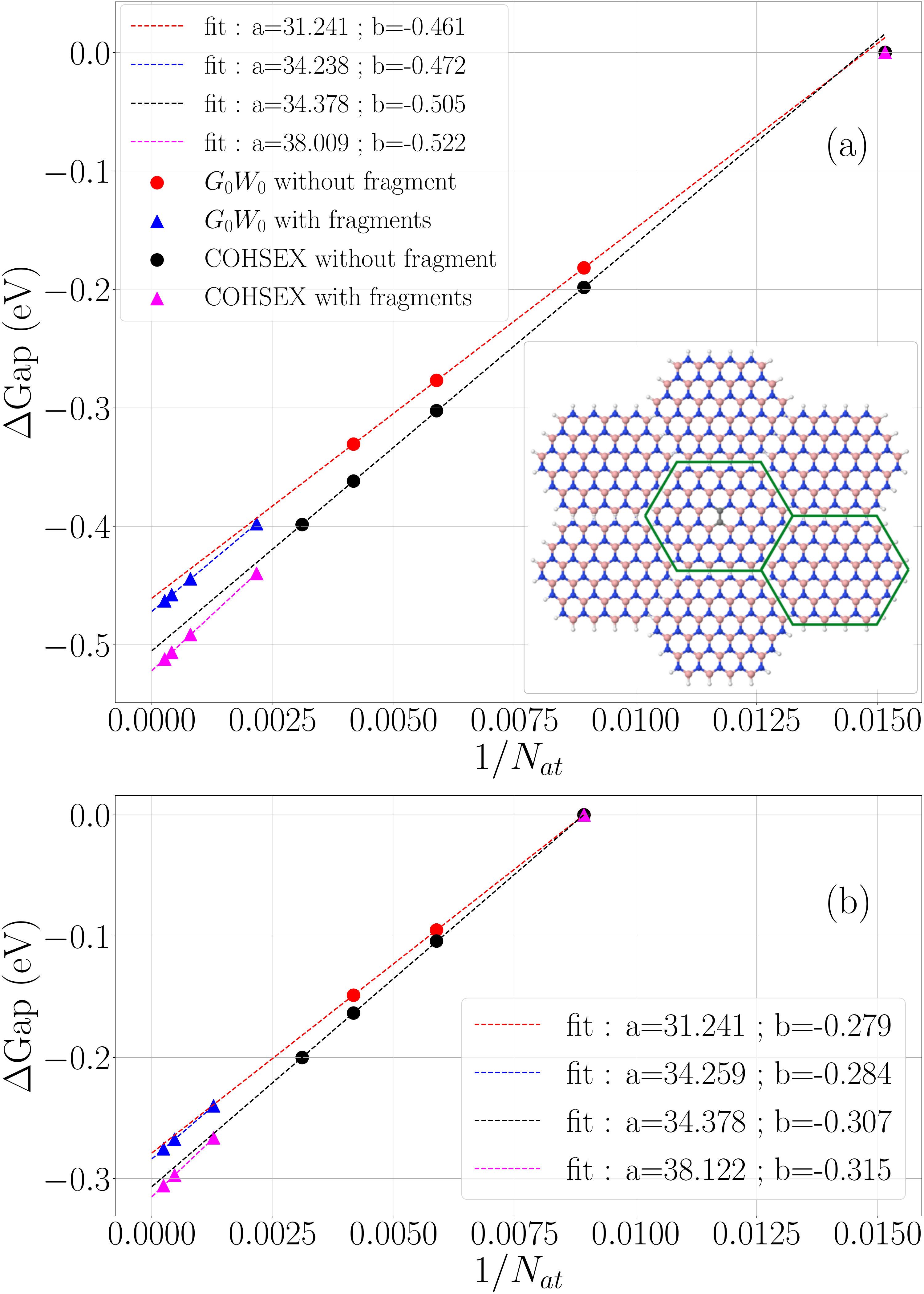}
	\centering
	\caption{ Evolution of the gap between the occupied and unoccupied CC defect energy levels with respect to (a) the CC@BN86 and (b) the CC@BN138 cluster gaps as a function of  1/$N_{at}.$ 
	The red (black) dots are full $G_0W_0$ (COHSEX) calculations for the CC@BN138, CC@BN202 and CC@BN278  clusters.
	The blue (pink) up-triangles represent  fragment $G_0W_0$ (COHSEX) calculations for (a) a central defected CC@BN86 cluster surrounded by up to four shells of undefected BN86 fragments, and (b) a central  defected CC@BN138 cluster surrounded by up to three shells of undefected BN138 fragments.  Dashed lines are $[a/N_{at} +b]$ fits.  (Inset) Symbolic representation of the fragment approach for building the susceptibility. The central CC@BN86 defected fragment is surrounded by undefected pristine BN86 first-nearest-neighbor fragments. Larger systems can be reconstructed by adding  2nd, 3rd, etc. neighbor fragments (not represented here). }
	\label{fig:monolayergap}
\end{figure}

In a previous study \cite{Winter_2021}, the monolayer  $CC$ dimer defect was studied using the $GW$ and Bethe-Salpeter formalisms within a finite-size cluster approach, following the variation of the quasiparticle gap and  optical excitation energies as a function of cluster size. It was shown in particular that the defect  electronic energy levels converge  very slowly with system size following a 1/R$^2$ scaling law, with $R$ the average radius of the defected \textit{h}-BN cluster considered. Such a scaling law  was shown to originate  from long-range polarization, or screening,  effects in a 2D system \cite{DAvino_2016}.  To reach such an asymptotic regime, allowing to extrapolate to the infinite size limit, $GW$ calculations on systems with increasing sizes well above two hundred atoms had to be considered. 
Such results are reproduced in Fig.~\ref{fig:monolayergap}(a) at the present $G_0W_0$@PBEh(0.4)   (red circles) and static COHSEX@PBEh(0.4)  (black circles) levels. The evolution of the gap of a CC defect at the center of BN138, BN202 and BN278 flakes (see geometries in SM~\cite{supplemental}) are given, taking as a reference the smallest CC@BN86 system. The dashed lines with corresponding colors are a $(1/N_{at})$ linear fit of the data, with $N_{at}$ the number of N/B atoms that is proportional to the squared radius  of the flake. We excluded the smaller CC@BN86 system for these fits to obtain an improved linear behavior, indicating that large systems must be considered to be in the limit where the fit is accurate.

The significant evolution of the gap  is the signature of the influence of enhanced screening by additional B/N atoms. In the asymptotic infinite monolayer size limit, the defect gap is closing by 0.46 eV ($G_0W_0$) or 0.51 eV (static COHSEX) as compared to the smaller CC@BN86 system. A small contribution from the closing of the gap originates from the evolution of the Kohn-Sham gap that can be  fitted with a $[1.87/N_{at} -0.05]$ eV law. This change in gap at the   DFT level is   an order of magnitude smaller than the evolution at the $G_0W_0$ level, indicating that long-range polarization effects are not captured at the Kohn-Sham DFT level. The closing of the gap due to polarization only, subtracting the ground-state Kohn-Sham contribution, amounts to $(P_e - P_h)=0.41$ eV at the $GW$ level and originates both from the stabilization of the defect occupied   level ($P_h$=0.20  eV) and of the unoccupied one  ($P_e$=-0.21 eV), where   the subscript h and e stand for hole and electron, respectively. 
We show now that the same results can be obtained by fragmenting the monolayer in domains reproducing the same screening, or polarization, effects on the central defect, but at a dramatically reduced cost allowing to reach much larger system sizes.

\begin{figure}[t] 
        \centering
	\includegraphics[width=8.6cm]{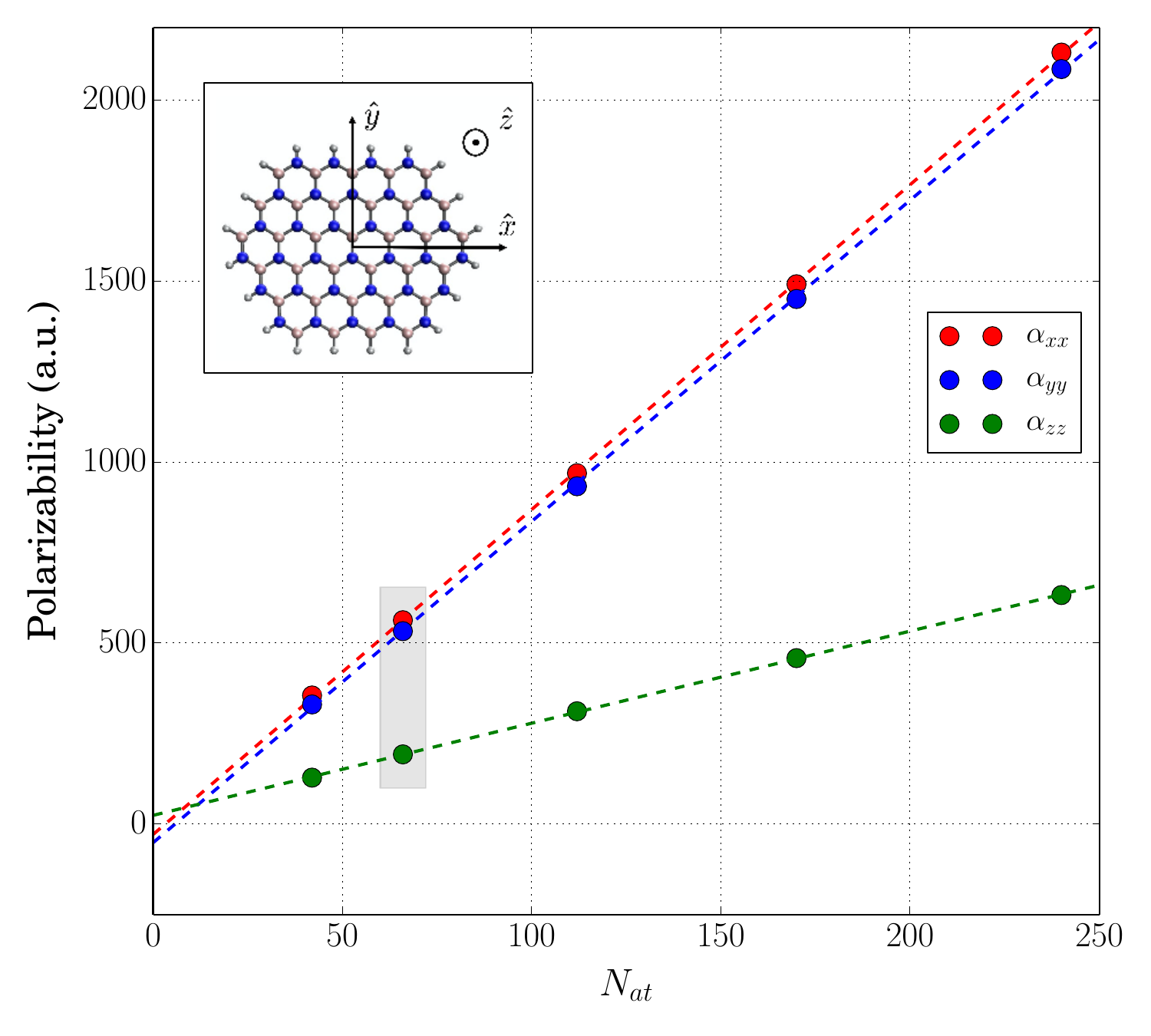} % for single-column, i.e. preprint
	\caption{  Evolution of the RPA static polarizability (in atomic units) of pristine \textit{h}-BN clusters with increasing number of B/N atoms. The three principal axes polarizabilities are given. The smallest contribution indicates the out-of-plane polarizability. The shaded data indicates the BN86 flake (see Inset)  used to fragment/reconstruct  extended monolayers in the following. }
	\label{fig:polar}
\end{figure}

To rationalize this strategy, we plot in Fig.~\ref{fig:polar} the (static)  polarizabilities of the pristine (undefected)  BN58, BN86, BN138, BN202 and BN278 flakes in the random phase approximation (RPA).  It is such an approximation that is used to build the screened Coulomb potential $W$ in standard $GW$ calculations (Eq.~\ref{eqn:Wscr}).
Such results show that the polarizability principal components follow accurately a linear relation with respect to the number of B/N atoms. As such, one can define a  polarizability-per-atom that is independent of the size of the considered cluster. We believe that such a behaviour is characteristic of insulating systems where charges cannot be displaced from one side of the system to another, and polarization proceeds rather by the creation of local  dipoles.   
%Note that the lines do not go strictly through zero, a small deviation that we interpret as the influence of the edge passivating hydrogen atoms. As shown below, the effect on the quasiparticle energies is in practice marginal.  

%%% FIG  3 %%%

The consequence of such a behaviour is that the effect on the central defect of polarizing a B/N atom located in an infinite monolayer, or at the same distance from the defect but in a finite size cluster, are very similar. 
%In the following, the defected or pristine BN86, BN138, etc. systems will also be labeled fragments, in the sens of pieces of an infinite monolayer. 
Such a finding allows to set up a fully \textit{ab initio} embedding scheme, with a central defected fragment embedded in rings of first, second, third, etc.  undefected neighbor fragments (see Inset Fig.~\ref{fig:monolayergap}a) reproducing the effect of a continuous layer at a dramatically reduced cost. As emphasized in the Methodology section, instead of calculating the independent-electron susceptibility $\chi_0$ of a system with increasing size, with a quartic scaling evolution in terms of the number of operations to perform, this susceptibility matrix  can be  calculated fragment by fragment, in a block diagonal fashion. 
%%The inversion of the Dyson equation to obtain $W$ - and the memory needed to store the related 2-body operators (susceptibilities, bare Coulomb potential, etc.) - limits the size of the number of fragments that can be studied. 
%\textcolor{red}{ For sake of indication, $GW$ and COHSEX calculations  for a monolayer of 57 BN86 fragments, namely up to 4th nearest neighbour around the central defected fragment, can be achieved with a typical total time (summing over all core contributions) of XXX   and XXX hours, respectively.   }

\begin{figure}[t] 
        \centering
	\includegraphics[width=8cm]{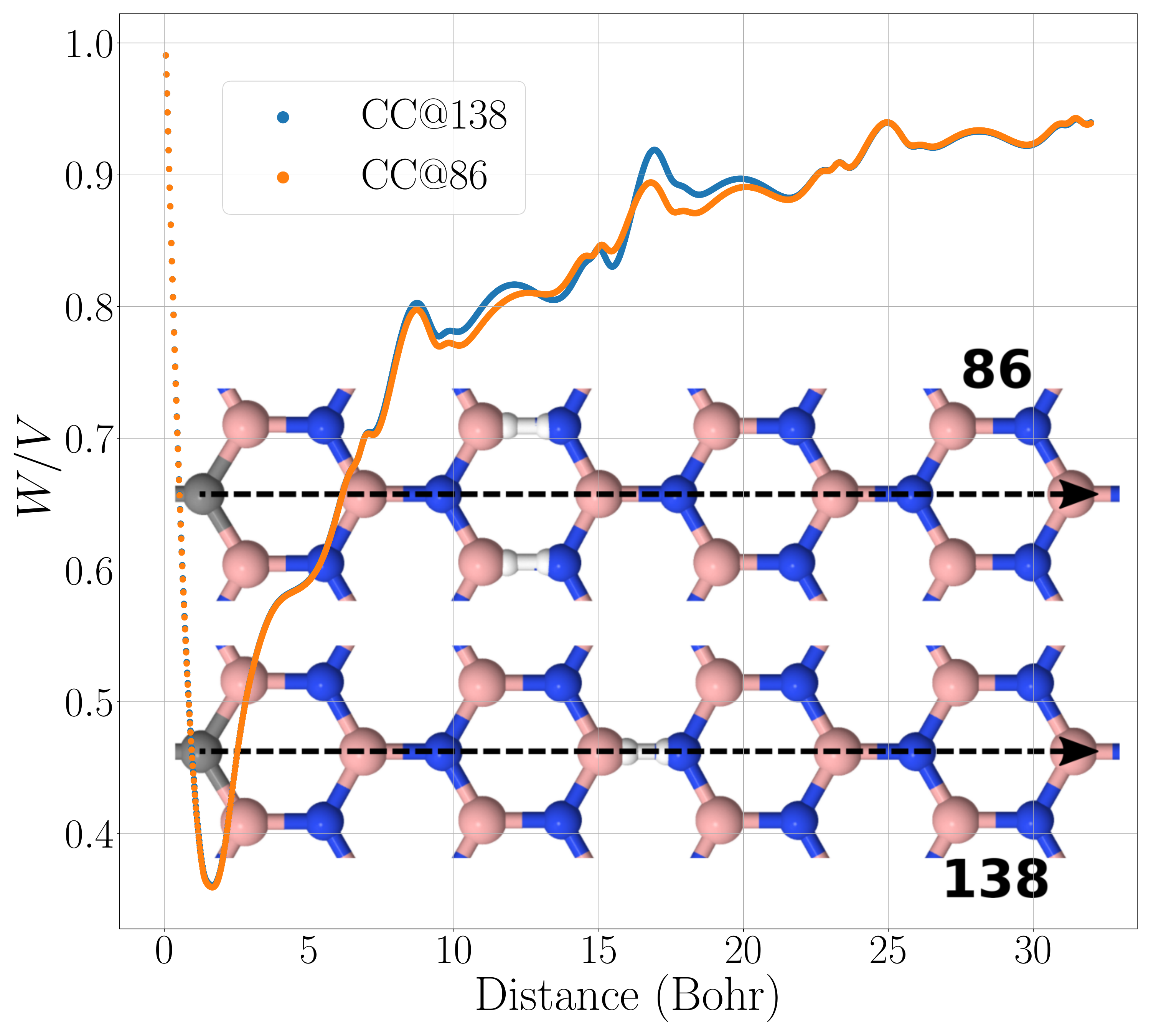} % for single-column, i.e. preprint
	\caption{   Ratio of the statically screened Coulomb potential $W({\bf r}; {\bf r}' ; \omega= 0)$ divided by the
bare Coulomb potential V with ${\bf r}$ fixed at the center of the CC bond and ${\bf r}'$ varying along the direction indicated by the horizontal black dashed lines.  
	The case of a defected monolayer reconstructed with BN86 fragments (orange plot) or  larger BN138 fragments (blue plot) are compared.  }
	\label{fig:WonV}
\end{figure}

To further confirm the  preservation of long-range screening upon fragmenting the monolayer, we plot in Fig.~\ref{fig:WonV} the ratio  of the statically screened Coulomb potential $W( {\bf r},{\bf r'}; \omega=0)$  over the bare Coulomb potential $V$, with $W$  obtained from Eq.~\ref{eqn:Wscr}   and the susceptibility $\chi_0$ built in a block-diagonal fashion from gas phase  $\chi_0^{defect}$ and $\chi_0^{pristine}$ fragment  susceptibilities. The ${\bf r}$ point is fixed to the CC bond center while ${\bf r}'$ varies radially in a specific direction  that goes close to H atoms when reforming the monolayer with BN86 fragments, and across H atoms when reconstructing the monolayer with larger BN138 fragments (see horizontal black dashed lines). The  $W/V$ ratio is very similar in both cases, with small differences when crossing or coming close to H atoms. In the vicinity of the CC bond, where the defect   states are localized, the $W/V$ ratio for the two systems are indistinguishable. As such,  changing the fragments size, that is changing the ratio of H atoms to B/N ones, hardly affects the screening properties. As expected for a 2D system, the $W/V$ ratio converges to unity in the long range limit \cite{Cudazzo_2010,Cudazzo_2011,Huser_2013,Latini_2015}, at odds with  the $1/\epsilon_M$ limit in 3D systems,  where $\epsilon_M$ is the macroscopic dielectric contant. This is a clear indication that long-range screening properties in 2D or 3D systems  cannot be reproduced within DFT by the same functional. A plot of the $W/V$ ratio in several directions is provided in the SM~\cite{supplemental}.

%%% FIG  3bis %%%

We now compare the $G_0W_0$ and COHSEX calculations performed for the CC@BNX systems, with X varying from 86 to 278, 
 to the fragment approach  where the monolayer is fragmented in BN86 defected (center) and undefected fragments with interacting polarizabilities. 
 %This is described in more details in the Supplemental Material where we show in particular that the presence of hydrogen atoms on the edge of each fragment does not lead to any difficulties when reforming the susceptibility of the   BN layer. 
Since the fragment approach can only account for polarization effects at the $GW$/COHSEX level, the small $[1.87/N_{at} -0.05]$ eV (see above) evolution of the Kohn-Sham gap with system size beyond the 86-atoms system is added perturbatively to the polarization correction. As emphasized above, such a perturbative approach is justified by the order-of-magnitude difference between gap changes at the Kohn-Sham and many-body levels.  As a result, extrapolations from the full (un-fragmented) calculations (red and black doted lines) and from the much cheaper fragment approach (up-triangles and corresponding fit)  differ   by no more than  11 meV (17 meV) in the $G_0W_0$ (COHSEX) calculations. As can be seen from the graph, the system  including the 4-th shell of nearest neighbor fragments, amounting to 57 flakes of 86 atoms each,  (see e.g. blue up-triangle at the $G_0W_0$ level) is already very close to the asymptotic infinite limit.

The very same exercise can be performed by reconstructing the infinite monolayer with larger BN138 fragments. Adopting larger fragments allows reducing the possible effects of edge H atoms polarizability. The results are represented in Fig.~\ref{fig:monolayergap}(b). Since we start from a larger system, the infinite size correction is smaller. The extrapolated value using the fragment approach falls within 5 meV (8 meV) of the extrapolated value obtained without any fragment approximation  at the $G_0W_0$ (COHSEX) level. As such, fragmenting the monolayer with small BN86 units or larger 138 units leads to very close results, confirming that  edge effects (e.g. hydrogen polarization) are negligible. 

 For further validation, we explored even larger systems using an efficient classical induced-dipole model of polarizable points \cite{Davino_2014,DAvino_2016}  with site polarizabilities fitted to reproduce the results of Fig.~\ref{fig:polar}. Such a microelectrostatic model allows to reach systems more than one order of magnitude larger than the one we can afford at the \textit{ab initio} level with the fragment approach. The outcome of these model calculations is that extrapolating the evolution of the gap to the infinite monolayer  on the basis of data acquired with much larger systems does not change by more than a meV the extrapolated value obtained with systems of the size we study \textit{ab initio} in the present fragments scheme. This is reported in more details in the SM~\cite{supplemental}.    

%Taking now the complete basis set limit HOMO-LUMO gap (see SM) for the C2@BN86 system, that amounts to X.XX eV, one obtains a $G0W0$@PBEh(0.4) defect gap of X.XX eV in the infinite size limit.

%The possibility to fragment monolayers in subunits reproducing faithfully long-range polarization effects is an important result allowing to reach the dilute defect limit at the many-body level with a dramatically reduced cost. Even though not the subject of the present study, such a finding further opens the door to studying systems where periodic boundary calculations are uneasy, such as systems composed of domains with different composition, having in mind e.g. the studies devoted to hybrid BCN planar systems \cite{Bernardi_2012a}, twisted layers with Moir\'{e} patterns \cite{Alexeev_2019}, or interfaces between 2D layers and disordered organic systems \cite{Bernardi_2012b} or water \cite{Kavokine_2022}.  We will be taking further advantage of that ``divide-and-conquer" strategy here below when exploring much larger multilayer systems. 

Coming back to the distinction between static and dynamical $GW$ calculations, we observe at that stage that the static COHSEX approximation  leads to polarization energies that are overestimated by about 10$\%$ as compared to $G_0W_0$ calculations. Namely, the static COHSEX approximation  slightly overestimates the closing of the gap due to enhanced dielectric screening.

\section{From the monolayer to   multilayers}

\begin{figure}[t]
	\includegraphics[width=8.6cm]{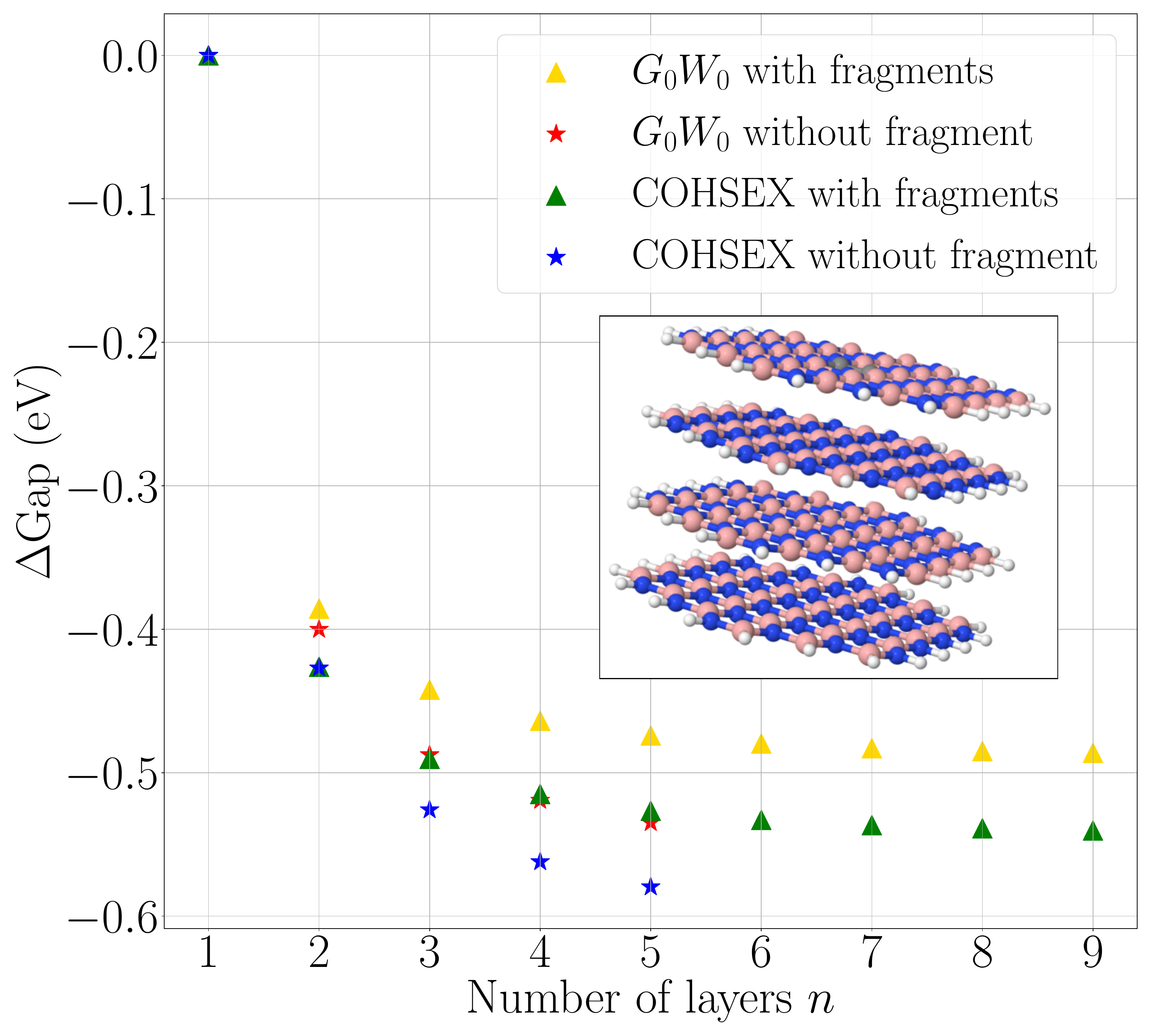}
	\centering
	\caption{ Evolution as a function of the number of layers of the defect gap of a defected CC@BN86 cluster located at the surface of a stack of BN86 undefected fragments in an AA' geometry. 
	The reference is taken to be the monolayer CC@BN86 fragment. Full $G_0W_0$ (red stars) and COHSEX (blue stars) are compared to fragment $G_0W_0$ ( yellow up-triangles) and COHSEX (green up-triangles). The wording ``fragments" means here a decoupling of layer wavefunctions. Inset: Schematic representation of the ($n$=4) system. Energies are in eV.	}
	\label{fig:stacksurf86}
\end{figure}

We now study the effect of layering on the defect energy levels, considering first   the carbon-dimer defect. We start our exploration by providing in Fig.~\ref{fig:stacksurf86}   the evolution of the defect $G_0W_0$ (red stars) and COHSEX     (blue stars) gaps from the monolayer  to   $n$-layer systems in a 86-atoms finite size layered geometry.  Namely, we here do not stack infinite layers but create a "cylinder" of stacked 86-atoms flakes, with a surface  defected CC@BN86 cluster deposited on top of undefected BN86 clusters in an AA' stacking fashion with 3.33~\AA~separation between layers (see schematic Inset Fig.~\ref{fig:stacksurf86}). We start with full $G_0W_0$ and COHSEX calculations, without any fragment approximation, stopping at 5-layers, representing already 430 atoms. 
% Performing such reference full blown $GW$ calculations on such systems  represents a very sizable computational effort while being, as shown below, very far from convergence in terms of system size. 

%%% FIG  4  %%%

As expected, interlayer screening reduces the  gap associated with the defect. At the $G_0W_0$ level in this 86-atoms stack geometry, the gap closes by as much as 0.4 eV upon adding a second layer, 0.49 eV with a third layer, etc.  Such a correction can be compared to the  0.106 eV closing of the gap at the Kohn-Sham level originating from the addition of a second layer, the difference between the pentalayer and the bilayer being reduced to 4 meV. This illustrates again that Kohn-Sham DFT cannot reproduce long-range screening unless a strategy is adopted to  readjust the functional for each number of layers  in order  to mimic increased screening with increasing number of layers. \cite{Smart_2018} Concerning many-body approaches,   we observe again that  the static COHSEX approximation overestimates the effect of screening, even though the error ranges from 30 to 50 meV, to be compared to the 0.4-0.6 eV total correction, namely an error not larger than 10$\%$. 

In a second step, we adopt a fragment approach, namely decoupling the wavefunctions at the Kohn-Sham level between adjacent layers, constructing the independent-electron susceptibility  in a block-diagonal fashion from the susceptibility of the isolated layers.  Such an approach has been used in several previous studies devoted to exploring the properties of stacked 2D systems (intercalating e.g. \textit{h}-BN, graphene, dichalcogenides)  \cite{Andersen_2015,Xuan_2019}. 
%In a first step, we consider as a same fragment the defected layer and its nearest neighbour layer. 
 As done in the previous Section for reaching the infinite   monolayer limit, the small Kohn-Sham shift between monolayer and multilayer systems is added perturbatively to the fragment $GW$ and COHSEX calculations to allow comparison with the full (unfragmented) calculations. 

 The present decoupling scheme is shown in Fig.~\ref{fig:stacksurf86} to underestimate the impact of screening on the defect gap beyond the bilayer system (compare the yellow up-triangles to the red stars at the $G_0W_0$ level and the green up-triangles to the blue stars at the COHSEX level). The error is however of the order of 50 meV at most, to be compared to a total correction of 0.4-0.5 eV. This relatively small error certainly confirms the success of the fragment approach in the 2D-genomics studies. 
 
 We further observe that this error is comparable in magnitude with that induced by the static COHSEX approximation, but with an opposite sign. As a result, and up to the 5-layers system, for which full reference $G_0W_0$ calculations were possible, there is a significant cancellation of errors between the static COHSEX approximation, that overestimates the effect of screening, and the fragment approximation that underestimates it. 
 %Obtaining reference (un-fragmented) many-body calculations for e.g. the 6-layers stack is becoming prohibitively expensive. 
% \textcolor{red}{We observe however that in the fragment approach, the error induced by the static COHSEX approximation does not increase with respect to the number of layers beyond n=3.  %% The improved accuracy of the static COHSEX approximation in the long-range limit was already justified in previous studies ([Neaton], Appendix Ref.~\citenum{}). }
We conclude from these tests that the fragment plus static COHSEX approximation reproduces with a   limited error full $G_0W_0$ calculations, thanks to a cancellation of error between the static and the fragment approximations. This cancellation reduces the 30-50 meV error associated with each separate approximation to a significantly lower value. While this cancellation is certainly fortuitous, since both errors are of different nature, we adopt this scheme in the following.  

%%% FIG  5  %%%
\begin{figure}[t]
	\includegraphics[width=8cm]{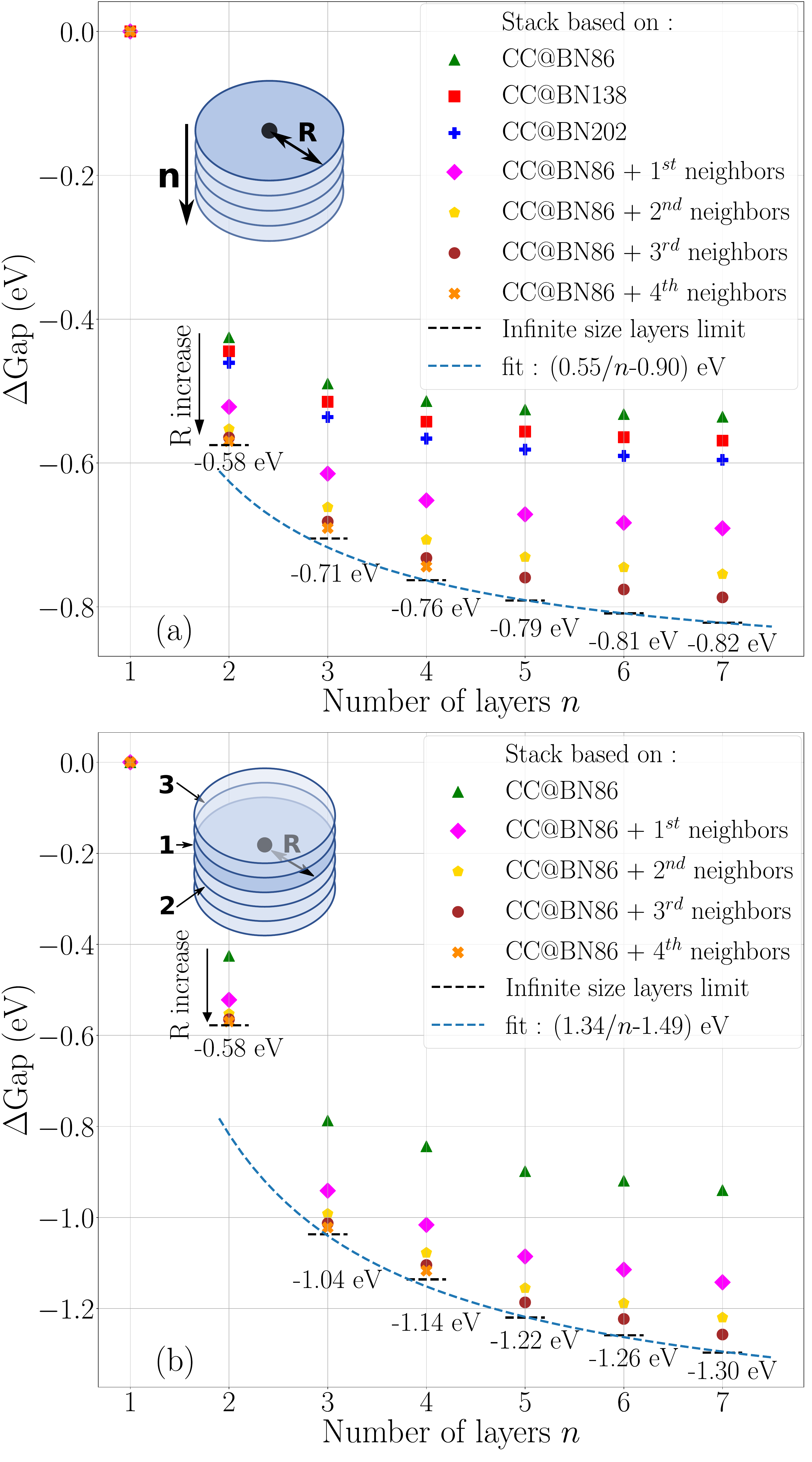}
	\centering
	\caption{ Evolution of the defect  gap (with respect to the monolayer) as a function of the number of layers (n) and the size of each layer. In (a) the defect is on the surface layer, while in (b) it is in the bulk with layers added alternatively on each side of the defect.  The horizontal dashed black lines are extrapolations to the infinite layers size for a given (n) value. The extrapolated gap closing value is given underneath. Extrapolated data points can be nicely fitted by a [$0.55/n-0.90$]  (eV) functional form for the defect at the surface and  [$1.34/n-1.49$]  (eV) for the defect in the bulk. Data obtained at the fragment $\Delta$COHSEX level. 	}
	\label{fig:stacklargerflakes}
\end{figure}

We now explore within the fragment static COHSEX scheme the evolution of the correction upon increasing the size of the layers, namely trying to converge towards the stacking of infinite size layers. We first study in Fig.~\ref{fig:stacklargerflakes}(a) the effect of stacking 138-atoms (red squares) and 202-atoms (blue pluses) flakes beyond the 86-atoms (green up triangles) systems studied in Fig.~\ref{fig:stacksurf86}. The defect is here again localized on the surface layer. 
%With increasing size, the Kohn-Sham shift from the monolayer to the bilayer decreases slightly down to -90 meV in the large size limit.  
%% [Shift KS 86 bi de -0.107ev, 138 de -0.099 eV et 202 de -0.094 eV] 
Clearly, as expected, increasing the size of each layer leads to increasing the polarization energy. 

Enlarging further the size of each layer becomes prohibitively expensive. 
We thus adopt the strategy of Section~\ref{sec:fragmono}  where we increase the in-plane lateral size by patching undefected BN86-atoms fragments around each CC@BN86 (defected layer) or BN86 (undefected layers) central fragment. Adding first,  second, third and fourth nearest neighbor  flakes  allows   reaching much larger layers. These systems are large enough to extrapolate to infinite size layers  for each ($n$) value.   This is again carefully checked with the semi-empirical model of polarizable points  (see SM~\cite{supplemental}). Reaching the asymptotic regime is shown to be more difficult   upon increasing the number of layers, suggesting that convergence criteria are related to the aspect ratio of the systems. 

The  polarization energies extrapolated to the infinite layer size for each ($n$) value are represented by horizontal black dashed segments.
Analytic derivation shows that in the limit of infinite size layers, polarization energies should scale as ($1/n$) as a function of the number ($n$) of layers. 
This is what we observe in Fig.~\ref{fig:stacklargerflakes}  with  a fit of the $n$=4 to 7 layers extrapolated values by a $[0.55/n - 0.90]$ eV functional form. 
Turning now to the bulk case (see Fig.~\ref{fig:stacklargerflakes}b), adding additional layers alternatively on one side and the other of the defected layer, one obtains an asymptotic evolution of the gap from the monolayer to the bulk that scales as $[1.34/n - 1.49]$ eV.  Even though fitting the $n$=5 to 7 layers data, the functional form yields a gap closing by 1.04 eV and 1.16 eV for  $n$=3 and $n$=4, respectively, in close agreement with the  1.04 eV and 1.14 eV explicit values. This indicates that the asymptotic regime is already quite accurate for very few layers. The $n$=2 case, very far from the asymptotic limit, can be much better estimated by the ``surface" fit that yields  -0.63 eV, within 0.05 eV of the explicit $\Delta$Gap=-0.58 eV  value.
 This rapid recovery of the ($1/n$) asymptotic behavior of the polarization energy with respect to the number of layers was again confirmed with the  semi-empirical micro-electrostatic model presented in the SM~\cite{supplemental}.
 
As expected, the closing of the gap by 1.49  eV in the bulk limit as compared to the monolayer is significantly larger than the 0.90 eV value found for a defect at the surface. 
We observe   that   for ($n$=3) the polarization energy correction accounts for $\simeq$70$\%$ (1.04 eV instead of 1.49 eV) of the ($n \rightarrow \infty$) limit. This  means that the nearest layers contribute significantly to the screening but that the true bulk limit requires additional layers. Using the obtained functional form, 90$\%$ of the polarization energy, or closing of the gap, is obtained for 8 additional layers, that is 4 layers on each side of the defected layer.   We provide in the SM~\cite{supplemental} a study of defects located in a sub-surface and sub-sub-surface layer.
%, showing that the polarization energy associated with the sub-sub-surface location is already within less than 10$\%$ of the bulk limit. 

Reproducing for the $C_BV_N$ defect the same study performed here above for the carbon-dimer defect in the bulk limit (see SM~\cite{supplemental}), the closing of the gap from the monolayer to the bulk amounts to 1.45 eV, very close to the 1.49 eV value obtained for the CC defects.  These values   are  consistent with the 1.27 eV, 1.31 eV and 1.49 eV gap variations obtained for the $C_B$, $C_N$ and $C_BV_N$ defects  within  Kohn-Sham DFT  using  Koopman's like conditions to adjust the fraction of exact exchange upon changing the number of layers \cite{Smart_2018}.  Such an agreement is remarkable given that the two techniques are significantly different.  

%(see Table III, Ref.~\citenum{Smart_2018}) . In that previous study, periodic boundary DFT calculations were performed imposing a Koopman's like condition to adjust the fraction of exact exchange for each number of layers.   
%However, the drop too small up to the 3-layer systems and then "huge" drop of 1.12 eV between 3layers and bulk .. but 3lyer is one layer on each side ?] 
%While the $GW$ formalism reconstructs explicitly  the nonlocal dielectric response upon adding layers, changing the amount of   exact exchange as a function of the number  of layers allows mimicking the change in long-range screening in a Kohn-Sham DFT calculation.  }

We  observe thus that the energy shift from the monolayer to the bulk is very weakly defect-dependent. The Kohn-Sham energy shifts between the monolayer and the multilayer systems is expected to change from one defect to another, depending on the local chemistry (hybridization, ionicity, etc.) Considering e.g. the $C_BV_N$ defect, the PBEh(0.4) Kohn-Sham gap is shown to close by about 60 meV from the monolayer to the 3-layer system on the basis of the stack of BN86 fragments. This value  can be compared to the 0.19 eV closing in the carbon-dimer defect case. These two shifts are very different in percentage but small in absolute value. We explore in the following Section the stability of the much larger polarization energy induced by long-range screening from one defect to another.   

%Reproducing our fragment $\Delta$COHSEX calculations for the $C_BV_N$ defect, we obtained a 1.XX eV value for the closing of the gap between the monolayer and the bulk, in excellent agreement with the tuned Kohn-Sham study, and 0.xx eV larger than the value obtained for the planar carbon-dimer defect. A Figure similar to Fig.~\ref{fig:gapC2inbulk} is provided in the SM for the $C_BV_N$ defect. The polarization energy associated with the $C_BV_N$ defect is found to be slightly larger than the planar carbon-dimer defect. We will discuss this small difference in the next paragraph devoted to the universality of the polarization energy.
%% Specifically, the donor ionization energies (downward relative to CBM) decrease from 2.52 to 1.49 eV (Delta = 1.03) for CB and 3.25 to 2.18 eV for CBVN, while the acceptor ionization energie (upward relative to VBM) decrease from 2.28 to 1.73 eV for CN and 2.14 to 1.57 eV for CNVB, irrespective of atomic chemical potentials. 

\section{ Polarization energies universality }

We now compare the evolution with layers number of the  quasiparticle energy levels for several defects, focusing on the CC carbon-dimer defect in its most stable first-nearest-neighbor conformation and its ``CC-$\sqrt{7}$" 4th-nearest-neighbor geometry,  together with the much studied $C_BV_N$ defect.  As discussed above, the evolution of the energy levels can be partitioned into a Kohn-Sham correction and the long-range polarization energies. While the Kohn-Sham correction has been shown to be short-range, involving mainly nearest-neighbor layers interaction, and rather small in magnitude,  we now show that  the larger polarization energy  is very much universal, namely system independent. 

\begin{figure}[t] 
	\includegraphics[width=8.5cm]{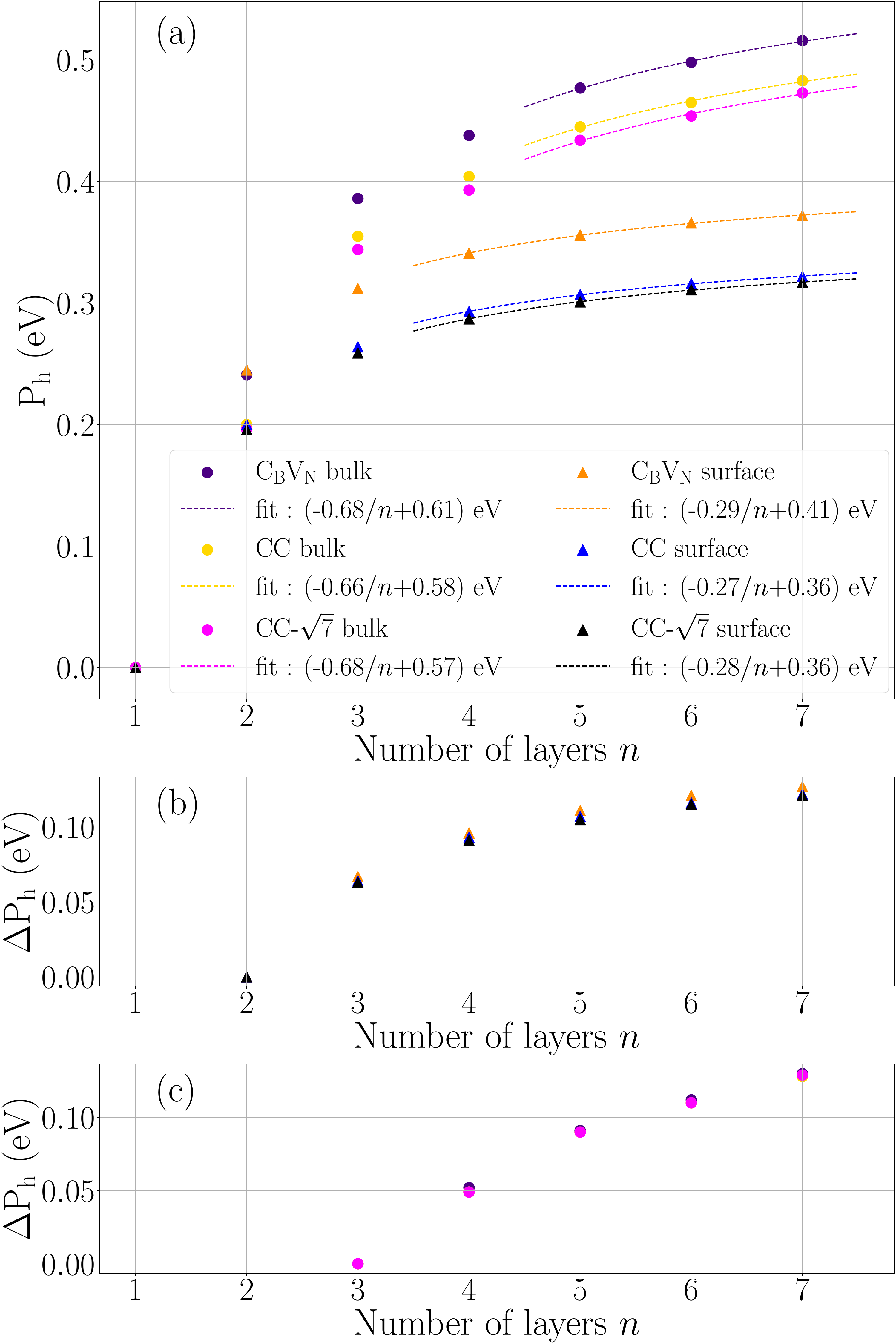}
	\centering
	\caption{ Polarization energy as a function of the number of layers for the $CC$, CC-$\sqrt{7}$ and $C_BV_N$ defect-associated  occupied  level  in the surface and bulk limits.  In (a) the polarization energy $P_h$ is taken with respect to the monolayer, while in (b) it is taken with respect to the bilayer for surface defects and in (c) with respect to the trilayer (one layer on each side of the defected layer) for the bulk ones.  The asymptotic fits are provided.	}
	\label{fig:universality}
\end{figure}

We plot in Fig.~\ref{fig:universality}(a) the evolution, from the monolayer to the $n$-layer case, of the defect occupied level polarization energy ($P_h$), considering both  cases of the defected layer in the bulk or at the surface. 
In the bulk limit, the polarization energies from one defect to another range from 0.57 eV to 0.61 eV, namely a variation around the mean value of about 3$\%$. Further, the prefactors governing the ($1/n$) evolution are remarkably close, comparing the evolution to the bulk limit and separately the evolution to the surface limit. The $C_BV_N$ defect is associated with the largest polarization energy. We attribute this to the $C_BV_N$ geometry with the carbon atom coming closer to its neighbor layer by $\sim$0.5~$\AA$, inducing a larger reaction field from this layer. The two other carbon-dimer defects are planar and, even though showing different in-plane extension and electronic properties, lead to very similar polarization energies. 

To further analyze the small differences from one defect to another, we provide in Fig.~\ref{fig:universality}(b) the evolution of the  polarization energy for the defect at the surface taking the bilayer system as reference, namely removing the evolution from the monolayer to the bilayer. Clearly, the various defects are characterized by residual polarization energies that are now within a very few meVs. In the case of the bulk, we reproduce the same exercise in Fig.~\ref{fig:universality}(c) but starting from the 3-layer case (one layer on each side of the defected one). Again, the evolution beyond the nearest-neighbor layers is very much defect independent. Such results confirm that the small  differences observed in Fig.~\ref{fig:universality}(a) between defects originate mainly from the response of the nearest-neighbour layers, the response of other layers being nearly completely independent of the defect chemical nature.   
 
%%% FIG  7  %%%

The present  0.57 eV to  0.61 eV evolution from the monolayer to the bulk of the defects occupied level  are consistent with the $\simeq$0.6 eV  stabilization energy of charged defects from the monolayer to the bulk as obtained in    DFT total energy calculations \cite{Wang_2020}.  In this previous  study, the   \textit{h}-BN substrate was replaced by a continuum model of dielectric function accounted for in the Poisson equation used to set up the Hartree potential. It is worth restating  here by comparison that the $GW$ formalism accounts  for polarization contributions to  electronic energy levels, not ground-state total energies as in Ref.~\citenum{Wang_2020}.  Performing the same $GW$ calculations for charged defects would require accounting for long-range screening effects both in the ground and excited states. This stands beyond the scope of the present paper.
%While Ref.~\citenum{Wang_2020} was concerned with the evolution upon adding layers  of charged defects   ground-state total energy, the present $GW$ formalism tackles charged  excitations associated with a photoemission experiment on neutral defects. In a ground-state DFT calculation, the stabilization of the added charge occurs mainly through the Hartree term, while in a $GW$ calculation, it is the exchange-correlation self-energy that conveys the screening response. In both cases, the dielectric environment stabilizes the added charge. 

We close   this section by concluding that  polarization energies, originating from environmental screening, are very much universal. Namely, they depend  on the host dielectric properties and on the defect location (surface or bulk), but hardly on the defect chemical nature. Indeed, even though the studied defects showed different chemical composition and band gap,   variations of the polarization energies from the monolayer to the bulk are within  a very few $\%$ of their mean value, and in practice even smaller if the defects do not break the planarity of the defected layer.  To better rationalize this universal behavior, we emphasize that the surrounding medium reacts to a localized added charge of which the exact spatial distribution, defect and level dependent, only affects its environment through higher order (dipolar, quadrupolar, etc.) contributions.  

The small observed differences  originate  nearly entirely from the effect of the nearest-neighbor layers. Further, the universal ($1/n$) long-range behavior, with $n$ the number of layers,  starts being valid for a very small number of added layers. In practice, this really means that while Kohn-Sham calculations  may have to be performed for a specific defect in the  bilayer and trilayer geometries  to capture  short-range ground-state crystal field  and hybridization effects, long-range polarization effects beyond the monolayer case can be accurately accounted for on the basis  of the simple scaling laws provided in this paper.   As such, expensive $GW$ calculations beyond the monolayer limit can be spared when exploring a large zoology of defects in their neutral state. 

\section{ Defect levels  with respect to boron-nitride valence band edge  }

All-electron finite-size Gaussian basis sets calculations provide energy levels directly related to the vacuum level. Locating defect levels with respect to the host semiconductor band edge may also be an important issue when it comes to discuss e.g. the evolution with the number of layers of the stability of defect charged states   \cite{Wang_2020}.  Contrary to localized states, addressing the energy of extended Bloch states from a finite size cluster calculation is a difficult task as compared to periodic boundary calculations. The flakes to be considered for extrapolating to infinity are much larger than a standard unit cell of the pristine semiconductor,  and the physical interpretation in terms of $k$-points is lost. However, the calculations performed above  allow in fact discussing the evolution of defect energy levels with respect to the \textit{h}-BN valence band edge.

\begin{figure}[t] 
 	\includegraphics[width=8cm]{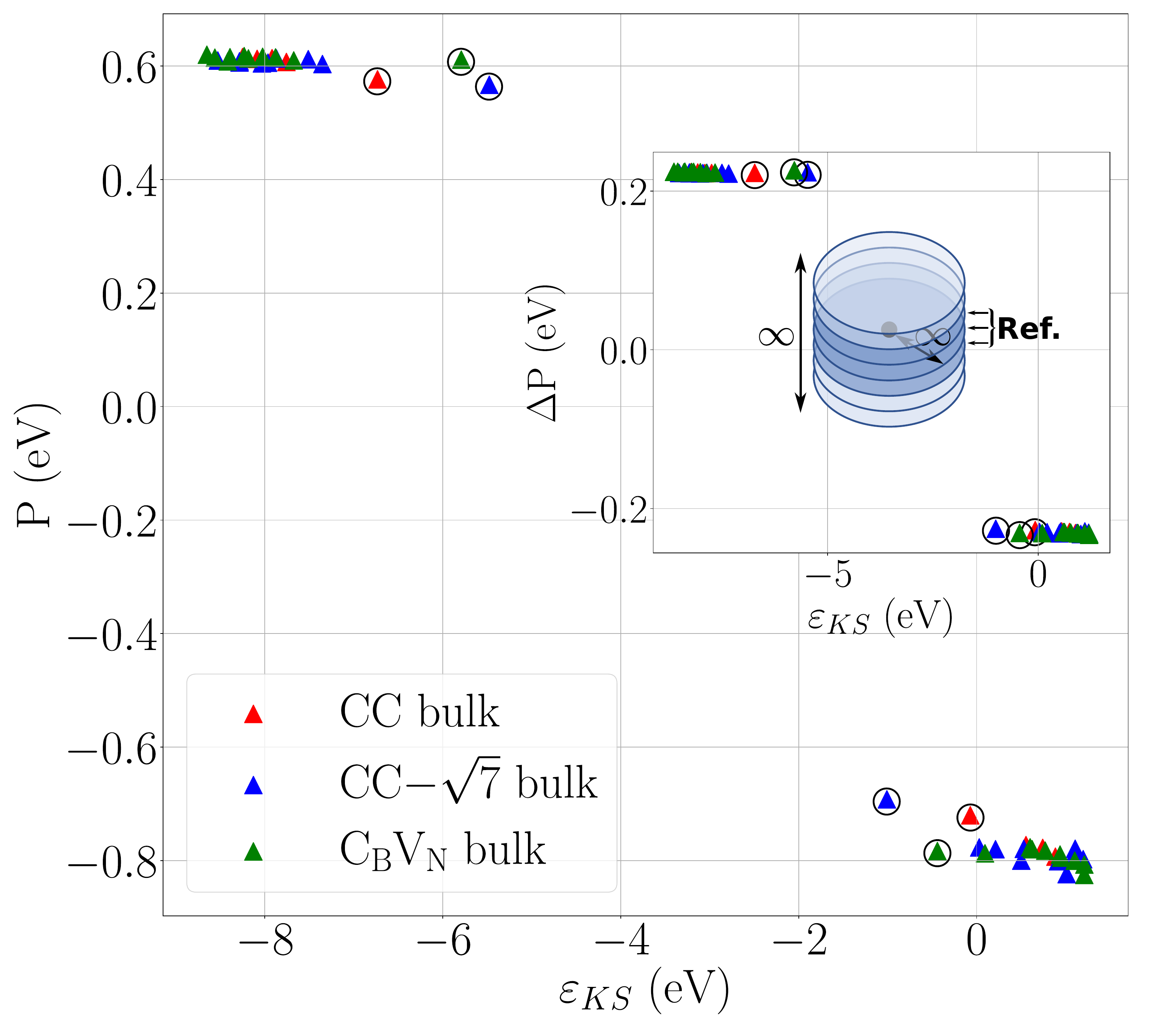}
	\centering
	\caption{ Polarization energies from the monolayer to the bulk for a larger set of occupied and unoccupied states, including defect   and \textit{h}-BN delocalized states, as a function of Kohn-Sham energy with respect to the vacuum level. 
	Defect states are circled. Most unoccupied states are unbound. (Inset) Evolution of the polarization from the 3-layer system (one undefected layer around the defected one) to the bulk.  }
	\label{fig:polar_all_states}
\end{figure}

In our defected flake calculations, the levels lying below the defect occupied level  are clearly  delocalized \textit{h}-BN states.  As such, the scheme presented here above can be used as well to extrapolate to the bulk limit not only the defects localized states, but further the highest occupied \textit{h}-BN states representative of the \textit{h}-BN top of the  valence band.   The results are represented in Fig.~\ref{fig:polar_all_states} which reports the bulk polarization energy, as compared to the monolayer,  for a large set of states around the gap in the case of the CC, CC-$\sqrt{7}$ and $C_BV_N$ defects. The states are ordered according to their Kohn-Sham energy. The defect states are circled. 

The salient result is that   polarization energies for the occupied states at the top of the \textit{h}-BN occupied states manifold are very stable around ${P_e}$$\sim$0.6 eV, to be compared to the 0.57-0.61 eV shift obtained for the localized defect levels (see Fig.~\ref{fig:universality}).  As such, screening by additional layers hardly affects the difference of energy between localized defect levels and delocalized \textit{h}-BN states at the top of the valence bands.  Additional analysis in the SM~\cite{supplemental}, considering larger fragments, confirms that the polarization energy is hardly dependent on the in-plane  localization of the considered states. This leaves hybridization and electrostatic effects in the ground-state as the only source able to modify significantly the position of the occupied defect levels with respect to the \textit{h}-BN valence band maximum (VBM). Here again, polarization energies, difficult to obtain at the Kohn-Sham DFT level,  are very much universal. Defect-dependent shifts with respect both to the vacuum level, but also the valence band maximum, can be obtained at the much cheaper (as compared to $GW$) Kohn-Sham DFT level.

%%% FIG  7 %%%

% [ This $P_h = 0.6$ eV polarization energy is comparable to the $\simeq$0.51 eV value reconstructed from the $G_0W_0$PBE data of Ref.~\citenum{Wang_2020} for the h-BN valence band maximum  (VBM)  subtracting tentatively the found -0.06 eV evolution of the VBM at the Kohn-Sham DFT level to the 0.45 eV upward shift of the $G0W0$@PBE VBM. ]
%% KS  CC-BN86    HOMO   LUMO            CC-BN138 
%%  mono    -6.7327   -0.0706          -6.7388      -0.1047
%%  tri     -6.5427  -0.0690           -6.5609     -0.1028
%%  Delta   +0.19      0.00             0.18        0.00
%% KS  CBVN-BN86    HOMO   LUMO          CBVN-BN138 
%%  mono         -5.7405   -0.4841       
%%  tri          -5.6318   -0.4323
%%  Delta         +0.11    +52 meV

As discussed above, the effect of stacking layers moves the defect occupied level towards the vacuum level (reduced ionization potential) by about 0.18 eV and 0.11 eV, respectively, at the Kohn-Sham ground-state level  for the CC and $C_BV_N$ defects. This correction is nearly entirely due to the interaction with nearest layers. Concerning \textit{h}-BN, it has been shown and analyzed in several papers that the top of the valence band at $K$ is very weakly dispersive in the AA' stacking configuration \cite{Ribeiro_2011,Wickramaratne_2018,Paleari_2018}. Namely, from the monolayer to the bulk, the top of the valence band remains  degenerate, dispersion growing slowly away from the $K$-point. Taking the vacuum level as a reference, the top of the valence band at $K$ was shown to go down in energy by about 0.1 eV \cite{Wickramaratne_2018} at the Kohn-Sham level with the HSE hybrid functional, demonstrating both weak electrostatic and hybridization effects. A similar value of 0.06 eV was found in Ref.~\citenum{Wang_2020} at the PBE level, confirming the stability of the valence band maximum at the Kohn-Sham ground-state level, namely in the absence of polarization effects between the monolayer and the bulk. Adding the -0.1 eV shift of the \textit{h}-BN VBM and the +0.1 eV to +0.2 eV shift upward of the occupied defect levels in the $CC$ and $C_BV_N$ cases, one finds that the energy spacing between the occupied defect levels and the VBM increases by a limited 0.2 eV to 0.3 eV from the monolayer to the bulk limit, an effect mostly related to electrostatics and hybridization effects accounted for at the Kohn-Sham level. 
%% \textcolor{red}{ Such values are very consistent with the periodic boundary DFT calculations \cite{Wang_2020}  ... well, check because really unclear ... }

We finally  address the challenging case of unoccupied states. Focusing first on the defect states, we observe in Fig.~\ref{fig:polar_all_states} that their polarization energy $P_e$ is larger (in absolute value) as compared to that associated with occupied defect levels. 
Unoccupied defect states are approaching the vacuum level, being weakly bound to the atomic layer. As such, the associated wavefunctions starts delocalizing away from the atomic layer   (see the analysis in the SM~\cite{supplemental}), inducing an enhanced polarization response from neighboring layers. Considering the effect of adding screening layers beyond the nearest-neighbor layers (see Inset Fig.~\ref{fig:polar_all_states}), namely calculating the polarization energy from the 3-layer system to the bulk, one observe that the effect of screening  becomes much more state-independent and symmetric between holes and electrons.   This is an indication that delocalization away from the defected layer remains small, 
at least for states close to the conduction edge, affecting mostly the response of the nearest layers but not beyond. This is similar to what was observed for the $C_BV_N$ defect with the out-of-plane C atom polarizing more strongly (as compared to the in-plane CC defect) the nearest layers. This overscreening effect due to loosely bound charges spilling out the BN layer  vanishes quickly for the response of layers located farther away.  As a result, polarization effects stabilize unoccupied defect levels by an energy $| P_e | \sim 0.7-0.8$ eV, or an energy $|P_e-P_h| \sim 1.3-1.4$ eV as compared to the \textit{h}-BN valence band maximum.

We will not attempt here to discuss the position of the \textit{h}-BN conduction band minimum. As shown in Fig.~\ref{fig:polar_all_states}, states above the unoccupied defect levels lie above the vacuum with a positive Kohn-Sham energy. As such, delocalized nearly-free-electron states are expected to be present at the conduction band minimum of \textit{h}-BN as documented in early  studies on \textit{h}-BN \cite{Blase_1994,Paleari_2018}. 
The difficulties associated with describing unbound states with localized basis sets, and the much larger dispersion of \textit{h}-BN states upon stacking at the conduction band minimum \cite{Wickramaratne_2018,Paleari_2018}, are serious limitations for the present localized-basis  fragments calculations. We report the reader to previous periodic-boundary  $GW$ and quantum Monte Carlo studies of the evolution of the electronic properties of pristine \textit{h}-BN from the monolayer to the bulk \cite{Paleari_2018,Smart_2018,Hunt_2020,Chen_2021}.

\section{Conclusion}

We have introduced and validated a fragment $GW$ approach for the study of defects in \textit{h}-BN multilayer systems where individual layers are fragmented in domains with non-overlapping wave functions. The resulting interacting  susceptibility and screened Coulomb potential are shown to   accurately describe the screening properties of   infinite monolayers in the vicinity of the dopant, reproducing at the few meV level the extrapolation of quasiparticle energies in the infinite layer size limit.  Such a divide-and-conquer scheme allows to study at the many-body level systems containing thousands of atoms, dramatically facilitating the study of dilute defects in monolayer, few-layers, surface or bulk \textit{h}-BN systems.
 
  The success of the fragmentation scheme for the block-diagonalization of the independent-electron susceptibility is associated with the large ionic gap character of the \textit{h}-BN substrate. For such systems, the response to a perturbation proceeds by the creation of  very localized induced dipoles, rather than the displacement of electrons over large distances. This can be restated by saying that  the susceptibility $\chi_0({\bf r},{\bf r}'; \omega)$ is very short-ranged, with an exponential decay in real-space in the case of insulators \cite{decay}. Along that line, exploring the robustness of this approach for other 2D systems with smaller gap (e.g. dichalcogenides) would be an interesting direction. In the present case, we find that our results are rather insensitive to fragment size and shape, provided  that the fragments are large enough (around a hundred atoms in the present study). As an additional ingredient, passivation of edge atoms, by repelling edge states away from the host gap, reduces the spurious contribution of edge polarization as compared to the bulk one. We do not exclude more efficient passivation strategies.  
 
As a first application, we have studied the evolution of the paradigmatic carbon-dimer and $C_BV_N$ defects energy levels  from the monolayer to n-layer systems, including the surface and bulk limits. The polarization energy, namely the evolution of the $GW$ electronic energy levels as a function of  increased screening upon adding additional layers, is shown to follow a nearly-universal (${\Delta P }/ n + P_{\infty}$) law with the number -$n$- of layers, where the rate ${\Delta P }$ and asymptotic value $P_{\infty}$ depend on the defect location (bulk or surface) but hardly on the defect type. Such an approach  rationalizes the evolution of defect energy levels as a function of the number of layers and allows  to easily extrapolate data obtained for the monolayer or very-few layer systems to the bulk. 

 This universal behavior can be rationalized by emphasizing that the surrounding polarizable substrate  reacts basically to an added charge associated with the photoemission process used to measure electron addition or removal energies. As such, the exact spatial distribution of this added charge, that depends on the defect type and chosen energy level, only contributes to higher order (dipolar, quadrupolar, etc.) terms to the perturbation felt by the polarizable substrate. While such arguments allow to rationalize asymptotic behaviors, it is remarquable that they start applying to the reaction of second, and even first, nearest-neighbor layers. While the present scheme ideally applies to localized defect states, our results suggest that similar polarization energies can be obtained for extended \textit{h}-BN Bloch states, leaving aside conduction band edge states that start delocalizing away from the plane of atoms. Special care should probably be taken with defects inducing large elastic deformations, requiring potentially larger fragments.

Together with the fragmentation scheme, we have shown that the neglect of neighboring layer wave function overlap, a common approximation in the 2D-genomics approach, leads to underestimating screening effects by up to 10$\%$. This error is however very largely canceled by neglecting the frequency dependence of the dielectric response in the calculation of the quasiparticle energy difference between the monolayer and   $n$-layer systems.  This  approximation takes the form of the  well-known static Coulomb-Hole plus screened exchange (COHSEX) approximation in the $GW$ framework. This static $\Delta$COHSEX schemes for calculating the increase in polarization energy upon adding layers, leads to overestimating polarization energies, significantly reducing the error introduced by decoupling wave function overlaps between layers. 

We further explored a multipole-like expansion of the long-range susceptibility, with a dramatic reduction of the auxiliary basis size used to express  density variations in the long-range. Together with fragmentation, this approach also contributed significantly to reducing the cost of performing many-body calculations.  For sake of indication, COHSEX calculations on  $\sim$10$^5$ electrons, as encountered for   7-layer systems, were conducted with a typical cost of 1600 CPU hours in the present fragment approach. While done here in a somehow \textit{ad hoc} fashion by removing high-angular momentum channels in the auxiliary basis, such a scheme may certainly be  improved.

Besides offering an alternative to periodic-boundary-condition calculations for study of the electronic properties of defect states in the dilute limit, the present scheme may offer a path to the study at the many-body level of disordered systems, including the stacking of layers of different chemical nature as in the field of 2D-genomics, but further the study of rotated layers with Moir\'{e} patterns \cite{Alexeev_2019}, disordered monolayers such as \textit{h}-BCN with the formation of segregated \textit{h}-BN and graphene   domains \cite{Bernardi_2012a,Bernardi_2012b}, or disordered interfaces with organic systems  or wet environments,  with intriguing contribution of 2D substrate polarization modes to the interactions at the interface \cite{Kavokine_2022}.  

Finally, the present calculations of $GW$ energy levels, accounting properly for long-range polarization effects, open  the way to optical absorption calculations within the Bethe-Salpeter formalism, where  excitonic electron-hole interactions are also expected to be screened by added layers.

%%%%%%%%%%%%%%%%%%%%%%%%
%\acknowledgements 
\begin{acknowledgments}
 DA is indebted to ENS Paris-Saclay for his PhD fellowship. This work was performed using HPC resources from GENCI-IDRIS (Grant 2021-A0110910016).
 XB and ID acknowledge support from the French Agence Nationale de la Recherche (ANR) under contract ANR-20-CE29-0005.
\end{acknowledgments}
%%%%%%%%%%%%%%%%%%%%%%%%
%\section*{Data availability}
%%%%%%%%%%%%%%%%%%%%%%%%
%The data that support the findings of this study are available within the article  and its Supplementary Informations.

%%%%%%%%%%%%%%%%%%%%%%%%

 \noindent\rule{4cm}{0.4pt}
%%  \bibliography{xavbib}

 %merlin.mbs apsrev4-1.bst 2010-07-25 4.21a (PWD, AO, DPC) hacked
%Control: key (0)
%Control: author (8) initials jnrlst
%Control: editor formatted (1) identically to author
%Control: production of article title (-1) disabled
%Control: page (0) single
%Control: year (1) truncated
%Control: production of eprint (0) enabled
%

%%%%%%%%%%%%%%%%%%%%%%%
%% Add supp info in manuscript for arXiv

\newpage

\setcounter{equation}{0}
\setcounter{section}{0}
\setcounter{figure}{0}
\setcounter{table}{0}
\setcounter{page}{1}
\makeatletter

\renewcommand\thepage{S\arabic{page}}
\renewcommand{\theequation}{S\arabic{equation}}
\renewcommand{\thefigure}{S\arabic{figure}}
\renewcommand{\thetable}{S\arabic{table}}
\renewcommand{\bibnumfmt}[1]{[S#1]}

\begin{center}
 \textbf{  Supplemental Material for :  \\
Universal polarisation energies for defects in monolayer, surface and bulk hexagonal boron nitride : A  finite-size fragments $GW$ approach  }
\end{center}

\vskip 1cm
%%%%%%%%%%%%%%%%%%%%%%%%

We provide here below supplemental informations concerning (I) the relaxed structures studied in the main manuscript,  (II) the evolution with distance and for several directions of the ratio of the statically screened Coulomb potential over the bare Coulomb potential, (III) microelectrostatic studies of polarization energies on the basis of system sizes varying over several orders of magnitude,  (IV)  the evolution of the $C_BV_N$ quasiparticle gap as a function of the number of layers in the bulk geometry,   (V) the evolution of the $CC$  occupied level polarization energy  as a function of the number of layers with the defect located in the subsurface and subsubsurface, respectively, (VI) the evolution of the \textit{h}-BN states polarization energy as a function of the fragment size over which they are confined, and in (VII) a study of the delocalization away from the atomic plane of unoccupied states.

\section{ Relaxed structures    }

\begin{figure}[h]
      \includegraphics[width=12cm]{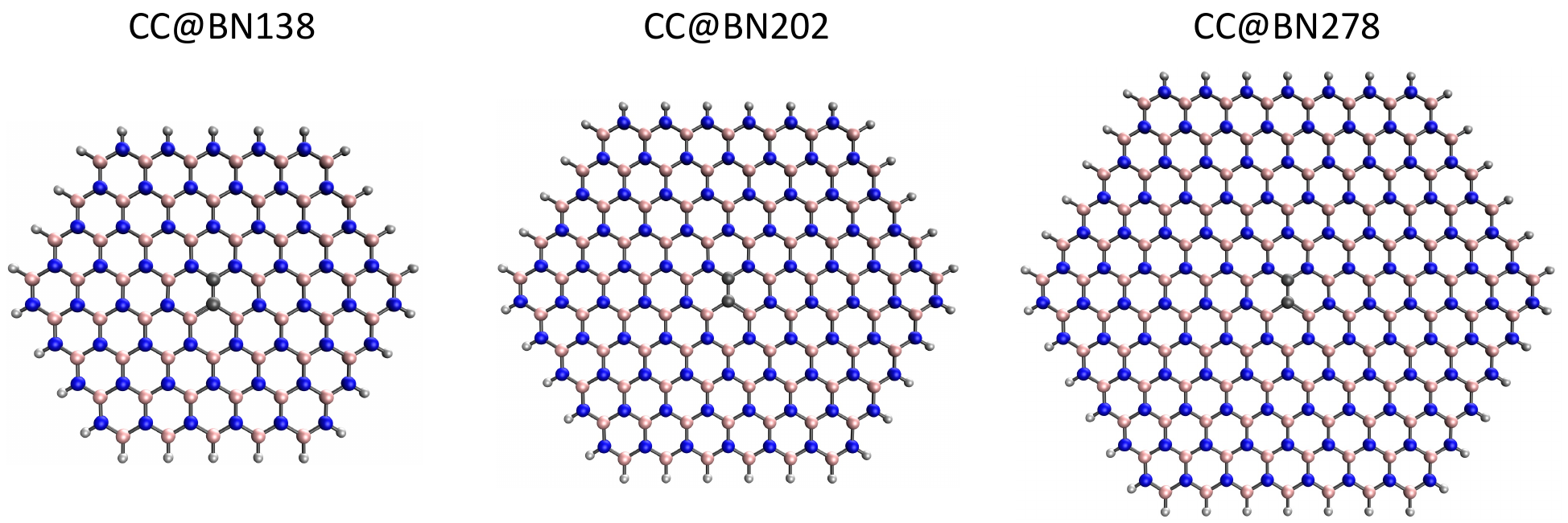} 
        \caption{ Representation of the C2@BN138, C2@BN202 and C2@BN278 structures. Hydrogen, boron, carbon and nitrogen atoms are in white, pink, black, and blue, respectively. Structures are not on scale. }  \label{fig:C2BNpict} 
\end{figure}

We provide in Fig.~\ref{fig:C2BNpict} a ball-and-stick representation of the carbon-dimer (CC) defect at the center of  BN flakes containing 138, 202 and 278 atoms. Edge atoms are passivated by hydrogens (in white).  Structures have been relaxed at the 6-311G(d) PBE0-D3 level.
Such systems are used to extrapolate to infinity the quasiparticle band gap of the CC dimer defect in the infinite size monolayer limit that is used as reference for the fragment approach (see Fig.~3 main manuscript).
\vskip .5cm

%%% FIG S1 %%%
 
\begin{figure}[h]
           \includegraphics[width=12cm]{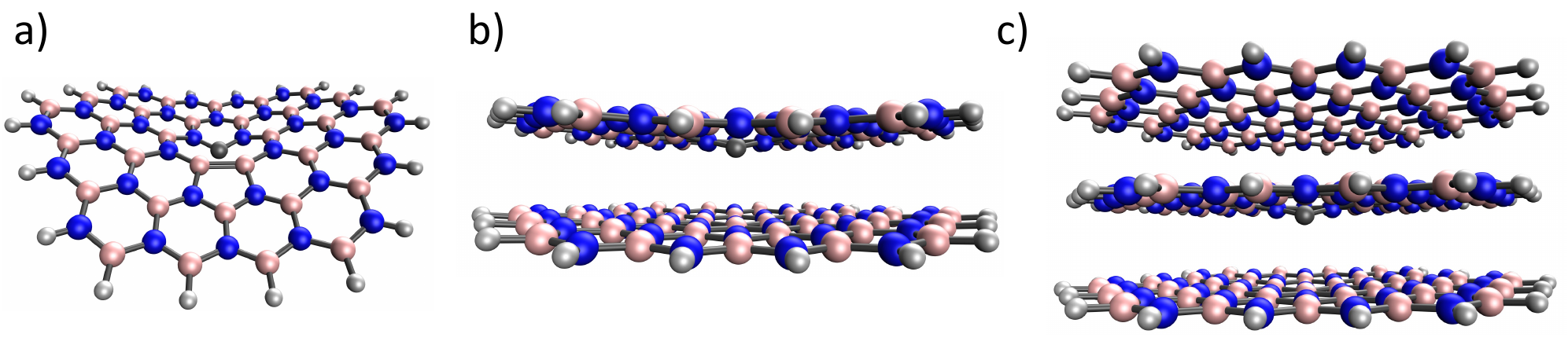} 
        \caption{ Representation of the $C_BV_N$ defect in (a) a BN86 monolayer, (b) a BN86 bilayer, and (c) a BN86 trilayer.  Structures are not on scale. }  \label{fig:CBVN} 
\end{figure}

We further represent in Fig.~\ref{fig:CBVN} the $C_BV_N$ defect at the center of a BN86 flake in a monolayer, bilayer and trilayer geometry (AA' stacking).  The relaxed monolayer displays a dramatically distorted non-planar structure, as a result presumably of the strain generated by the B-B bond across the nitrogen vacancy.
However, the interaction with a second layer [Fig.~\ref{fig:CBVN}(b)] restores planarity but with a C atom that goes off-plane by about 0.55~\AA. The corrugation of the defected layer, that is the maximum out-of-plane displacement of B/N atoms, is of the order of 0.25~\AA.
The addition of a 3rd layer [Fig.~\ref{fig:CBVN}(c)] leads to the same geometry, with the C atom that goes off-plane by about 0.52~\AA\ and a reduced corrugation of 0.22~\AA. 
In the main manuscript, when studying the evolution of the electronic energy levels from the monolayer to a $n$-layer system in the surface configuration (defect at the surface), the monolayer geometry is extracted from the bilayer one. 
Similarly, when studying the evolution of the electronic energy levels from the monolayer to a $n$-layer system in the bulk configuration, the monolayer geometry is extracted from the trilayer geometry. 
\vskip .5cm
 
%%%%% FIG S2 

 \section{ Screened Coulomb potential $W( {\bf r}, {\bf r}' ; \omega=0)$ along specific directions }

\begin{figure}[h]
               \includegraphics[width=7cm]{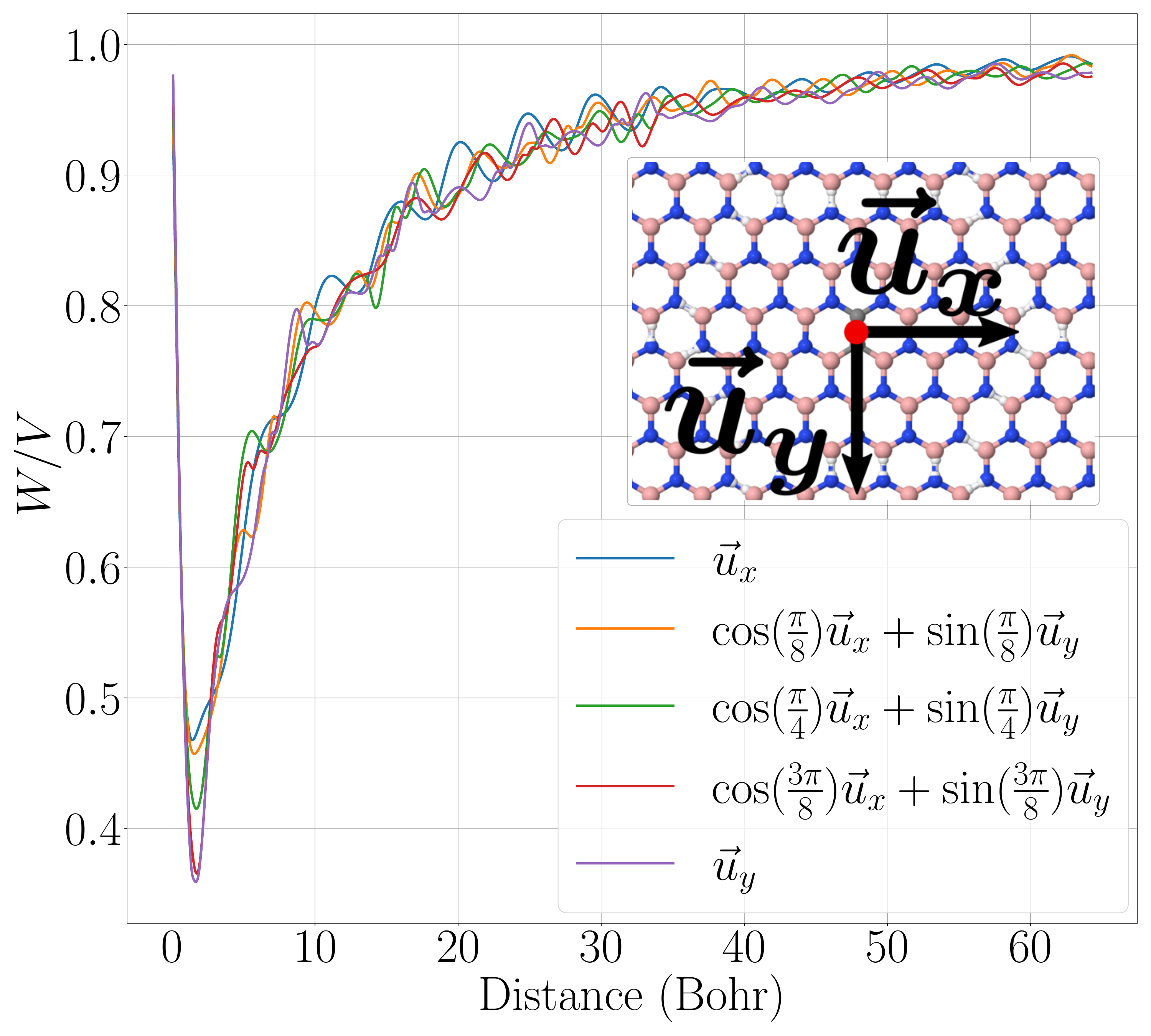}
        \caption{ Representation of the ratio of the static screened Coulomb potential $W( {\bf r}, {\bf r}' ; \omega=0)$  divided by the bare Coulomb potential $V$ along several directions for ${\bf r}'$ with $\bf r$ fixed to the center of the defect CC bond. 
         }  \label{fig:wplot} 
\end{figure}

We plot in Fig.~\ref{fig:wplot} the ratio of the statically screened Coulomb potential $W( {\bf r}, {\bf r}' ; \omega=0)$  divided by the bare Coulomb potential $V$ in the case of the carbon-dimer defect with ${\bf r}$ located at the center of the CC bond and ${\bf r}'$ displaced along specific directions. In the main text, the plot was provided along the $\vec{u}_y$ direction.
The small variations for a given $| {\bf r}-{\bf r}'|$ distance indicate  the magnitude of local field effects that fade away at large distance. 
 We observe that  there is no screening at short range  [$W/V$ goes to unity for   $| {\bf r}-{\bf r}' |  \rightarrow 0$] and that further  in the \textit{h}-BN monolayer there is  no screening in the long-range [$W/V$ goes to unity for   $| {\bf r}-{\bf r}' | \rightarrow +\infty$] as documented in several analytic studies \cite{Cudazzo_2010,Cudazzo_2011,Huser_2013,Latini_2015}.  

%%% Fig. S3

\section{ Microelectrostatic study of scaling laws in the very large system size limit }

To validate the infinite size extrapolations performed on the basis of \textit{ab initio} calculations, we adopt a microelectrostatic semi-empirical approach allowing to study systems   comprizing up to $\sim$78000 atoms/layer. 
Such studies confirm the accuracy of  extrapolations based on data acquired at the \textit{ab initio} level on systems comprizing a few thousand atoms. 
In brief, B/N atoms are replaced by  polarizable centers equiped with an atomic polarizability fitted to reproduce the RPA polarizability of the BNX (X=86, 138, 202, etc.) clusters (see  Fig.~4 of the main manuscript).
In the presence of the field generated by a charge associated with the CC@BN86 defect levels, as given by  \textit{ab initio} calculations, induced dipoles are generated on these polarizable centres. 
The energy of these dipoles, interacting with the source charge and with each other, is minimized. The reaction field of these equilibrated dipoles on the source  charges gives the polarization energy.
In such calculations, only the polarization energy, not the ground-state electrostatic and hybridization effects, are accounted for. 
Such a scheme is implemented in the MESCAL code  \cite{Davino_2014,DAvino_2016}.

%%%% Fig S4
\begin{figure}[t]
           \includegraphics[width=16cm]{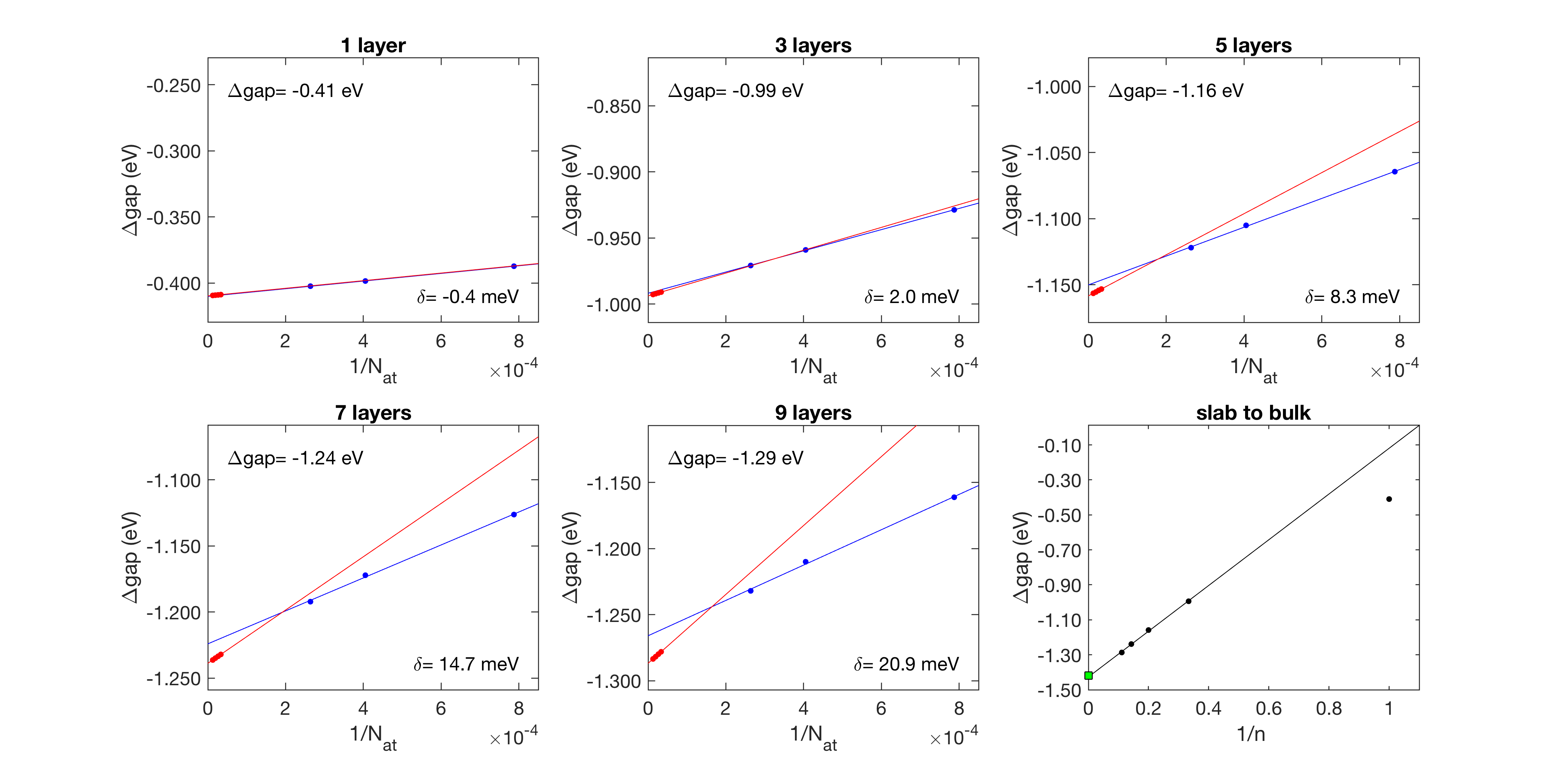}          
           \caption{ Evolution of the gap between the occupied and unoccupied CC defect energy levels with the system size, computed with a classical induced-dipole model of atomistic resolution for 2D slabs of a given thickness (up to $n=9$ layers, the defect is placed in the middle layer), and as a function of the number of layers (last panel). Slab calculations: blue points are results obtained on systems of roughly the same size (up to $N_{at}=3768$~atoms/layer) as those employed in the fragment COHSEX calculations in the manuscript (Figure 7); red points refer to results obtained on much larger systems (up to $N_{at}=78166$~atoms/layer); lines are linear regressions performed on the two data sets. The difference between the two intercepts $\delta $ (noted in the bottom-right corner of each plot) quantifies the error associated with the extrapolation on relatively-small systems.  This error is negligible for the monolayer and grows with the slab thickness reaching 20.9 meV for the 9-layer system. 
           The gap variation extrapolated in the infinite slab limit is noted in the top-left corner of each panel. The last panel shows the evolution of the gap variation as a function of the film thickness (black dot), converging linearly in $1/n$ to the bulk value (green square), the latter obtained upon extrapolating results for spheres of increasing radius. }  \label{fig:mescal} 
\end{figure}

We plot in Fig.~\ref{fig:mescal} the semi-empirical polarization energy contributing to the closing of the CC@BN86 defect-associated gap (namely $\Delta$gap= $P_e - P_h$) as a function of system size (number of polarizable centers $N_{at}$ per layer). 
Taking the case of the monolayer, the semi-empirical polarization energy leads to a difference in gap from the CC@BN86 fragment to the infinite monolayer of 0.41 eV, in close agreement with the \textit{ab initio} data
that predict a 0.46 eV  gap evolution (see Fig.~3 main manuscript) with a 0.05 eV contribution from the evolution of Kohn-Sham eigenvalues not accounted for in the semi-empirical approach. 
We compare the ($1/N_{at}$) extrapolation obtained on the basis of data points acquired for  system sizes equivalent to that used in the \textit{ab initio} calculations (blue dots and fits),
namely a central CC@BN86 cluster surrounded by up to 4th nearest-neighbour pristine BN86 fragments, to calculations and fits (in red) obtained with systems more than an order of magnitude larger.
In the monolayer and trilayer cases, the difference is within the meV, confirming that systems used at the \textit{ab initio} level are large enough to obtain a reliable extrapolation.
This difference becomes larger when the number of layer increases but remains within 15 meV up to the 7-layer systems studied in the main manuscript. It is a general result that convergence criteria rely on the aspect ratio (radius versus height) in the case of cylinders. Finally (lowest right "slab to bulk" plot), the ($1/n$) behavior with layer number $n$ of the polarization energy is confirmed using the microelectrostatic model. 
We observe that the ($n=3$) case is already on the ($1/n$) fit.

\section{ Monolayer to bulk quasiparticle gap evolution for  the $C_BV_N$ defect }

\begin{figure}[h]
      \includegraphics[width=7cm]{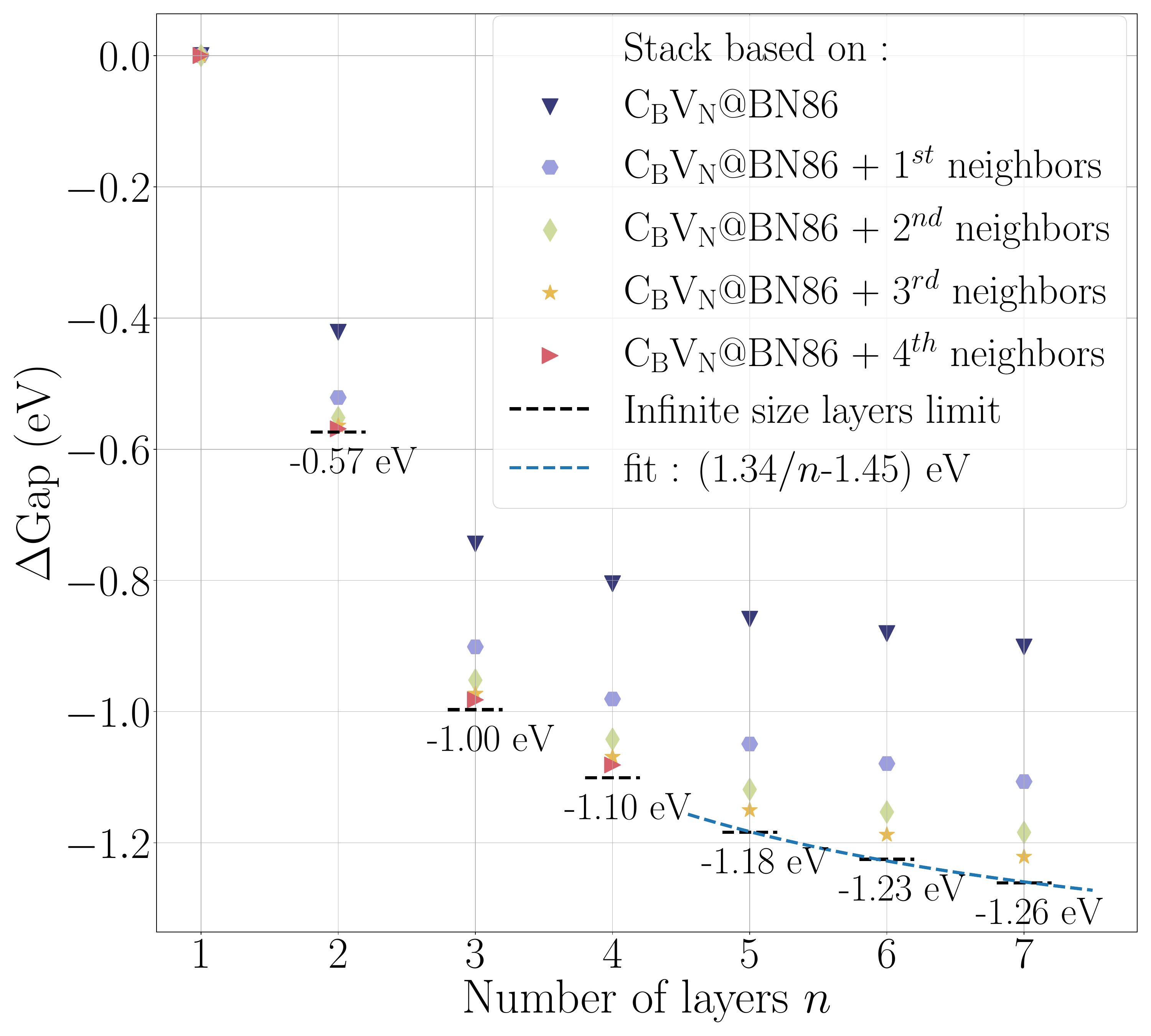}
        \caption{ Evolution of the quasiparticle gap as a function of the number of layers and as compared to the monolayer for the $C_BV_N$ defect in the bulk geometry (layers successively added on both side of the defect).
        }  \label{fig:cbvnbulk} 
\end{figure}

We present in Fig.~\ref{fig:cbvnbulk} the evolution of the $C_BV_N$ quasiparticle gap, taking as a reference the monolayer,   as a function of the number of layers for the $C_BV_N$ defect in the bulk limit.  Layers are  added on each side alternatively of the central defected layer, respecting AA' stacking. The defected layer geometry, with the C atom located $\sim$0.52~\AA\ off-plane, has been relaxed using a 3-layer geometry within the DFT 6-311G(d) PBE0+D3 approach. In the fragment approach, the evolution of the gap properly contains the -40 meV and -57 meV evolution (closing) at the Kohn-Sham level from the monolayer to the bi- and trilayers, respectively, additional layers hardly changing the Kohn-Sham gap value.  As such, both ground-state electrostatic and hybridization effects, together  with polarization (screening)  effects at the many-body level, are accounted for.

\section{ Polarization energy for defects in sub(sub)surface position }

\begin{figure}[h]
           \includegraphics[width=8.6cm]{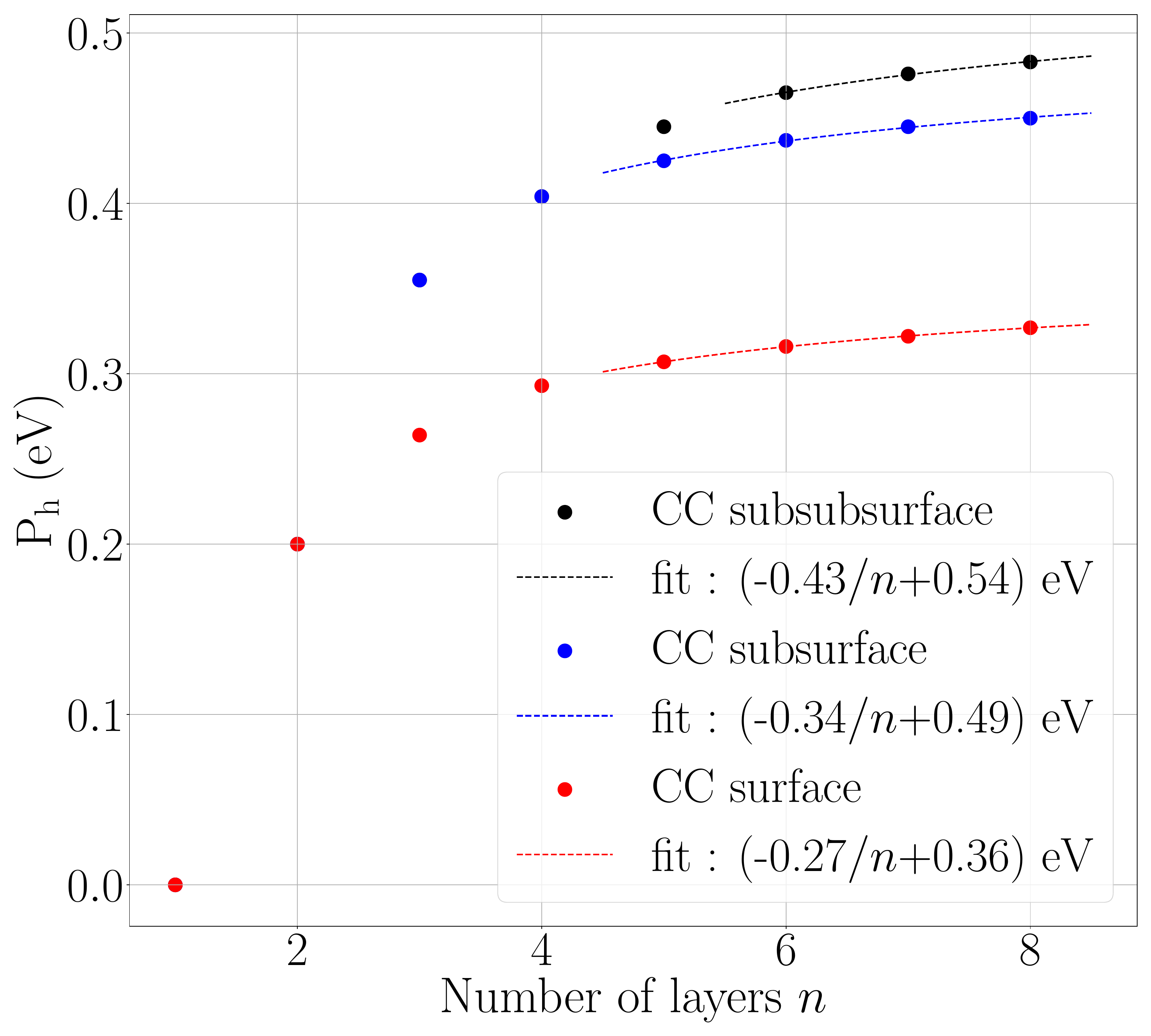}          
           \caption{  Polarization energy as compared to the monolayer for the CC dimer defect occupied level. Both the subsurface and subsubsurface positions are explored. 
           }  \label{fig:subsurface} 
\end{figure}

We complete our study by considering in Fig.~\ref{fig:subsurface} the evolution of the occupied defect level polarization energy for the  CC dimer defect from the monolayer to a $n$-layer system with the defect  in subsurface and subsubsurface positions. In the subsurface case, the (n=2) layer is added on one side of the defected layer, all layers for ($n>2$) being added on the other side.
 In the subsubsurface case, the layers are added alternatively on each side of the defected layer up to n=5 (2 layers on each side) and then added subsequently only on one side.   
The $P_{\infty}$=0.54 eV polarization energy for the subsubsurface position can be compared to $P_{\infty}$=0.58 eV for the bulk position (see Fig.~8a main manuscript). 
The surface case  is reproduced from  Fig.~8a main manuscript for sake of comparison.

\vskip .5cm

\section{ Polarization energy of \lowercase{\textit{h}}-BN states as a function of fragment size  }

\begin{figure}[t]
           \includegraphics[width=8.6cm]{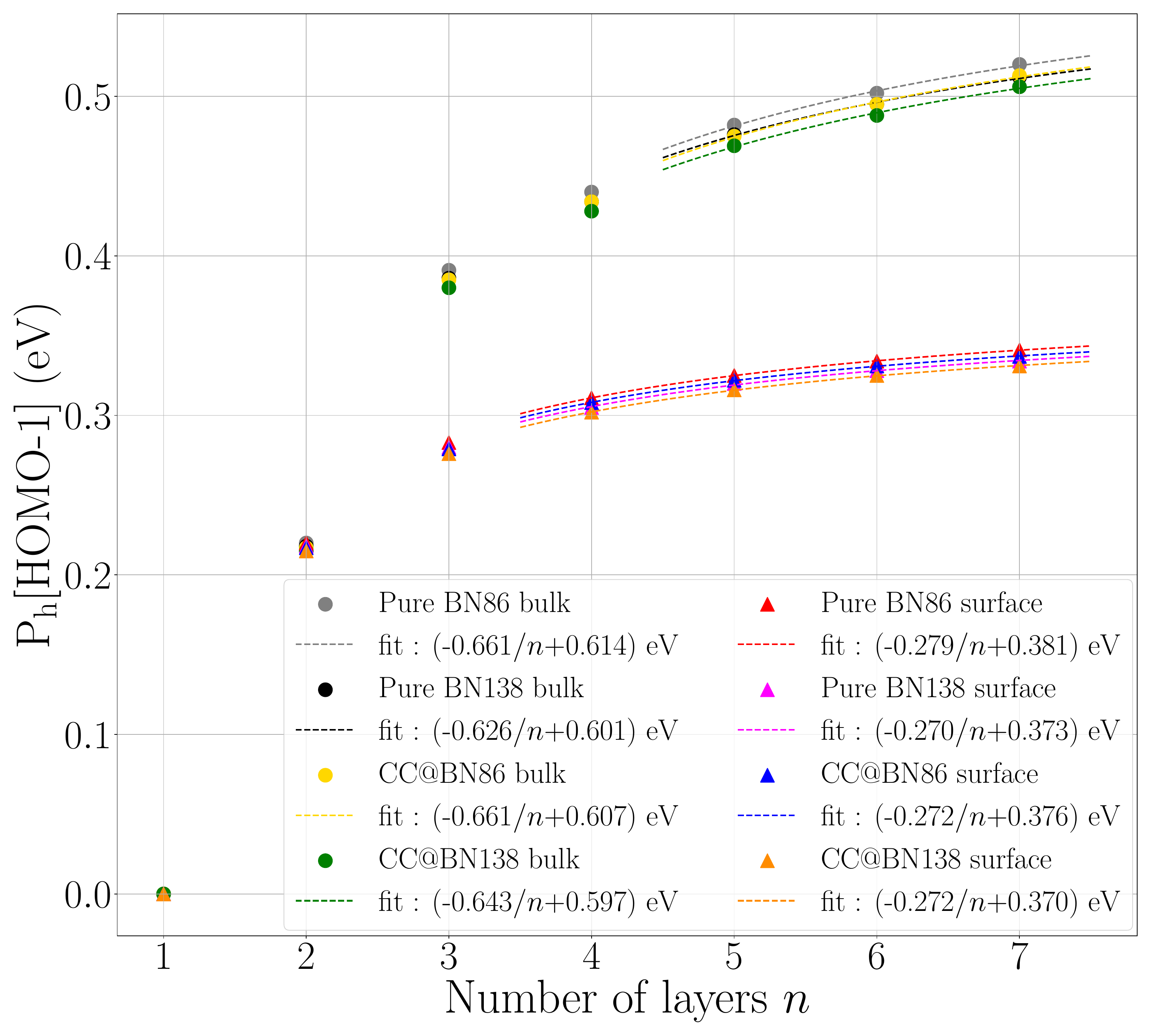}          
           \caption{  Polarization energies for the monolayer to a n-layer system  for the highest occupied \textit{h}-BN state delocalized over a central fragment with size BN86 or BN138.
           Both the highest occupied (HOMO) orbital of a pristine (pure) central BN86 or BN138 fragments, and the highest-but-one (HOMO-1) orbital of   defected CC@BN86 or CC@BN138 fragments, are studied in the bulk and surface limits.
           }  \label{fig:hbn138} 
\end{figure}

We test in the present Section the evolution of the polarization energy associated with the highest occupied \textit{h}-BN delocalized states as a function of the fragment size.
Since in our  approach the \textit{h}-BN  states are confined to the central fragment, varying the size of such a fragment, from BN86 to BN138, allows to assess the influence of the in-plane spatial extension.
In Fig.~\ref{fig:hbn138}, we compare the polarization energies obtained for the highest occupied (HOMO) orbital of  pure (undefected) central  BN86 and  BN138 clusters, surrounded by fragments of the same size in the same layer or in surrounding layers.
As in the main text, we also look at the evolution of the highest-but-one (HOMO-1) occupied orbital of defected  central CC@BN86 and CC@BN138 clusters. Both the surface and bulk limits are considered.
Variations of the ordered of 10 meV are observed, from $P_h \simeq 0.61$ eV to $P_h \simeq 0.60$ eV in the bulk limit, confirming the stability of the obtained polarization energies for extended \textit{h}-BN Bloch states as a function of fragment size.

%%% Fig S7

\section{ Localization of unoccupied versus occupied states }

\begin{figure}[t]
           \includegraphics[width=7cm]{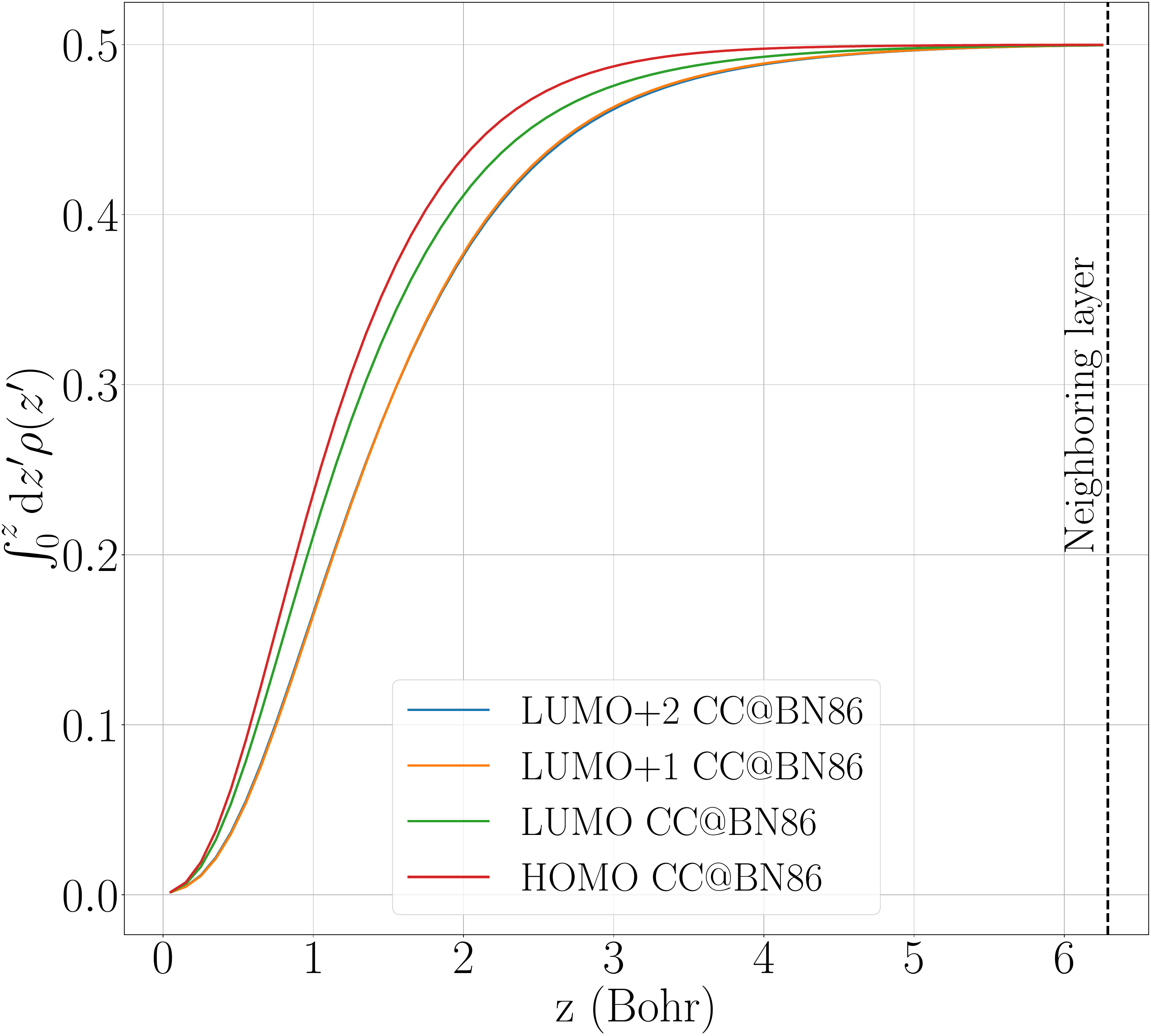}          
           \caption{  Integrated charge accumulated within a distance -z- of a CC@BN86 defected fragment for various states. 
           HOMO   indicates the highest  occupied orbital, while LUMO, (LUMO+1) and (LUMO+2) indicate the lowest,  lowest-plus-one and lowest-plus-two unoccupied ones.
           The vertical dashed line indicates the position of the neighboring layer. }  
           \label{fig:charge} 
\end{figure}

We plot in Fig.~\ref{fig:charge} the   charge density accumulated within a distance -z- of a CC@BN86  defected system for selected states. 
Only the charge on one side of the atomic layer is accounted for.
The graph evidences the delocalization away from the atomic plane with increasing Kohn-Sham energy.
This can be used as a mean to better understand the increase of polarization energies (in absolute value) for unoccupied states with increasing energies. 
As discussed in the main text, this delocalization affects mainly the polarization response from the nearest-neighbor layers, 
additional screening originating from layers added beyond being nearly independent of the chosen state, restoring the electron-hole polarization energy symmetry.

%%%%%%%%%%%%%%
%%%%%%%%%%%%%%
\end{document}